\documentclass[review]{elsarticle}

\usepackage{lineno,hyperref}
\usepackage{graphicx}
\usepackage{epstopdf}
\usepackage{subcaption}
\usepackage{color}
\usepackage{rotating}
\usepackage{bm}

\journal{Journal of \LaTeX\ Templates}

\biboptions{authoryear}

\begin{document}

\begin{frontmatter}

\title{Characterizing the feedback of magnetic field on the differential rotation of solar-like stars}

\author[label1]{J. Varela}
\ead{Jacobo.Varela@cea.fr}
\author[label1,label2]{A. Strugarek}
\author[label1]{A. S. Brun}

\address[label1]{AIM, CEA/CNRS/University of Paris 7, CEA-Saclay, 91191 Gif-sur-Yvette, France}
\address[label2]{D\'epartement de Physique, Universit\'e de Montr\'eal, C.P. 6128 Succ. Centre-Ville, Montr\'eal, QC, H3C-3J7, Canada}



\modulolinenumbers[5]

\begin{abstract}

The aim of this article is to study how the differential rotation of solar-like stars is influenced by rotation rate and mass in presence of magnetic fields generated by a convective dynamo. We use the ASH code to model the convective dynamo of solar-like stars at various rotation rates and masses, hence different effective Rossby numbers. We obtained models with either prograde (solar-like) or retrograde (anti-solar-like) differential rotation. The trends of differential rotation versus stellar rotation rate obtained for simulations including the effect of the magnetic field are weaker compared with hydro simulations ($\Delta \Omega \propto (\Omega/\Omega_{\odot})^{0.44}$ in the MHD case and $\Delta \Omega \propto (\Omega/\Omega_{\odot})^{0.89}$ in the hydro case), hence showing a better agreement with the observations. Analysis of angular momentum transport revealed that the simulations with retrograde and prograde differential rotation have opposite distribution of the viscous, turbulent Reynolds stresses and meridional circulation contributions. The thermal wind balance is achieved in the prograde cases. However, in retrograde cases Reynolds stresses are dominant for high latitudes and near the top of the convective layer. Baroclinic effects are stronger for faster rotating models.

\end{abstract}


\end{frontmatter}


\section{Introduction}

It is well known that there is a correlation between magnetic activity and rotation of stars \citep{1976ApJ...204..589D,1984ApJ...287..769N,2003AandA...397..147P}. Rapid rotators show a stronger and more intense magnetic activity \citep{1999ApJ...524..295S,2010Sci...329.1032G} than slower rotators such as the Sun for which the averaged magnetic field is weaker \citep{1981ApJ...248..279P}, therefore a detailed analysis of the differential rotation (DR) is mandatory to understand the magnetic activity of the stars \citep{1996ApJ...466..384D}. 

Dopper imaging \citep{1997MNRAS.291....1D,2005MNRAS.357L...1B}, asteroseismology \citep{2004SoPh..220..169G,2013AandA...560A...4R,2014AandA...572A..34G}, classical spot models \citep{2014AandA...564A..50L} and short-term Fourier-transform \citep{2014MNRAS.441.2744V} are methods to infer the differential rotation, while photometric and spectroscopic variability are good indicators of the magnetic activity along the star's activity cycle \citep{1995ApJ...438..269B,2009AandA...501..703O}. The combination of both sources of information helps to constrain the trends linking rotation with stellar differential rotation and magnetic activity, data available thanks to recent missions as CoRoT or \textit{Kepler}. Recent analysis revealed weak dependency between DR and star's rotation ($\Delta \Omega \propto \Omega^{0.15}$) \citep{2005MNRAS.357L...1B,2013AandA...560A...4R}), larger in case of star's temperature ($\Delta \Omega \propto T_{eff}^{8.92}$ \citep{2005MNRAS.357L...1B,2013AandA...560A...4R}) and $\Delta \Omega \propto T_{eff}^{8.6}$ \citep{2007AN....328.1030C}). The  differential rotation defined in these communications is $\Delta \Omega = \alpha \Omega_{eq}$ with $\Omega_{eq}$ the angular velocity at the equator and $\alpha$ the relative horizontal shear of the differential rotation between the equator and the pole. $\Omega_{eq}$ and $\alpha$ are deduced from the observations.

Several authors have performed global 3D magnetohydrodynamic (MHD) simulations to model differential rotation and stellar magnetism in the convection zone \citep{2006ApJ...641..618M,2010ApJ...715L.133G,0004-637X-735-1-46,2011AandA...531A.162K,2014AandA...570A..43K,2015ApJ...809..149A,2015AandA...576A..26K}, particularly for solar like stars \citep{2004ApJ...614.1073B,2010ApJ...711..424B,2011ApJ...731...69B,2011ApJ...742...79B}. These studies pointed out the large magnetic temporal variability and the critical effect of stellar rotation and mass on magnetic field generation through dynamo mechanism, leading in some parameter regimes to configuration with cyclic activity \citep{1983ApJS...53..243G,Gilman2,2013ApJ...762...73N,2013GApFD.107..244K,2013ApJ...777..153A,2015arXiv150704434G,2015ApJ...809..149A}. Several studies pointed out the effect of a stable region underneath the convection zone on the lengthening of the stellar dynamo cycle period \citep{2015arXiv150704434G,2015ApJ...813...95L}.

The present study is focused on solar-like stars, G and K stellar classes. Observations indicate that this group of stars show very different magnetic activity \citep{1990MmSAI..61..559S,1999ApJ...514..402P}, with short \cite{2010ApJ...723.1583M} and long cycles \citep{1995ApJ...438..269B}, consequence of the range of masses, rotation, differential rotation, age, effective temperature or metallicity measured \citep{1984ApJ...279..763N,2010ApJ...713L.169C,2014ApJ...790...12B,2014AandA...572A..34G,2041-8205-790-2-L23}. We analyze the correlation between differential rotation and magnetism in solar-like stars using the anelastic spherical harmonic code (ASH) \citep{2004ApJ...614.1073B}, performing several convective dynamo MHD simulations for star with different masses and rotation rates (Rossby numbers). One first achievement of the study was to simulate stars with prograde (solar-like) and retrograde (anti-solar-like, equator rotates slower than the poles) differential rotation \citep{2011AN....332..897M,2011AN....332.1045B,2014MNRAS.438L..76G,2015AandA...576A..26K}. The aim of this study is to analyze the effect of a magnetic field on the star's differential rotation \citep{2004SoPh..220..333B,2014ApJ...789L..19F}.We show that the trends of the differential rotation with the stellar mass and rotation for MHD simulations are in better agreement with the observational trends than equivalent hydro simulations \citep{1996ApJ...466..384D,2005MNRAS.357L...1B,2013AandA...560A...4R}.

The article structure is as follows: section 2; we describe the ASH code, the boundary and initial conditions of the different models as well as the key parameters of each simulation. Section 3; we study the large scale flows for the different models analyzing the time averaged kinetic and magnetic energy of the system, differential rotation as well as the trends of differential rotation versus star's rotation rate and mass obtained for hydro and MHD cases. Section 4; we analyze the angular momentum balance in the models studying the mean radial and latitudinal fluxes transport. Section 5; we study the baroclinity and thermal wind balance for typical prograde and retrograde cases. Section 6; conclusion, discussion and perspectives of present study.

\section{Numerical model}

In this section we present the main features of the ASH code, describing the boundary and initial conditions of the numerical model and our choice of the global parameters.

We perform 3D simulations of convective dynamo action that consist in solving the Lantz-Braginski-Roberts (LBR) form of the anelastic MHD equations for a conductive plasma in a rotating sphere \citep{2011Icar..216..120J}, a formulation that improves the energy conservation in stable stratified regions \citep{2012ApJ...756..109B,2013ApJ...773..169V}. The code ASH performs a large-eddy simulation that uses a pseudo-spectral method with the spherical harmonics expansion in the horizontal direction for the entropy ($S$), magnetic field ($\mathbf{B}$), pressure ($P$) and mass flux. The density ($\rho$), entropy, pressure and temperature ($T$) are linearized about the spherically symmetric background values, denoted by the symbol ( $\bar{}$ ). The solenoidality of the mass flux and magnetic vector fields is maintained by a streamfunction formalism \citep{2004ApJ...614.1073B}. The equations solved by ASH are \citep{2014AandA...565A..42A,2015ApJ...809..149A}:
$$ \boldsymbol{\nabla} \cdot \bar{\rho}\mathbf{v} = 0 $$
$$ \bar{\rho} \frac{\partial \mathbf{v}}{\partial t} = -\bar{\rho}\mathbf{v}\cdot\boldsymbol{\nabla}\mathbf{v} - \boldsymbol{\nabla}\bar{\omega} + \frac{Sg}{c_{p}}\mathbf{r} + 2\bar{\rho}\mathbf{v} \wedge \mathbf{\Omega}_{0} + \frac{1}{4\pi}(\boldsymbol{\nabla} \wedge \mathbf{B})\wedge\mathbf{B} + \boldsymbol{\nabla} \cdot D  $$
$$ \bar{\rho} \bar{T} \frac{\partial S}{\partial t} = \bar{\rho} \bar{T} \mathbf{v} \cdot \boldsymbol{\nabla}(\bar{S}+S) - \boldsymbol{\nabla} \cdot \mathbf{q} + \Phi $$
$$ \boldsymbol{\nabla} \cdot \mathbf{B} = 0$$
$$ \frac{\partial \mathbf{B}}{\partial t} = \boldsymbol{\nabla} \wedge [\mathbf{v} \wedge \mathbf{B} - \eta \boldsymbol{\nabla} \wedge \mathbf{B}] $$
with the velocity field $\mathbf{v} = v_{r}\mathbf{r} +  v_{\theta}\boldsymbol{\theta} +  v_{\varphi}\boldsymbol{\varphi}$, the magnetic field $\mathbf{B} = B_{r}\mathbf{r} +  B_{\theta}\boldsymbol{\theta} +  B_{\varphi}\boldsymbol{\varphi}$, the angular velocity in the of the rotation frame $\mathbf{\Omega}  = \Omega_{0} \mathbf{z} $, $\mathbf{z}$ the direction along the rotation axis, $g$ the magnitude of the gravitational acceleration and $\bar{\omega} = P / \bar{\rho}$ is the reduced pressure in the LBR implementation. The motions not resolved by the numerical mesh are parametrized as effective eddy diffusivities $\nu$, $\kappa$ and $\eta$ that account for the effect of the subgrid-scales transporting momentum, heat and magnetic field. The diffusion tensor $D$ and the dissipative term $\Phi$ are defined as:
$$ D_{ij} = 2\bar{\rho} \nu \left[ e_{ij} - \frac{1}{3}\boldsymbol{\nabla} \cdot \mathbf{v} \delta_{ij} \right]$$
$$ \Phi = 2\bar{\rho} \nu \left[ e_{ij} e_{ij} - \frac{1}{3} (\boldsymbol{\nabla} \cdot \mathbf{v})^{2} \right] + \frac{4\pi\eta}{c^{2}}\mathbf{J}^{2}$$
with $e_{ij}$ the stress tensor and $\mathbf{J} = c/4\pi\boldsymbol{\nabla}\wedge\mathbf{B}$ the current density. The energy flux $\mathbf{q}$ has a radiation flux and an inhomogeneous turbulent entropy diffusion flux:
$$ \mathbf{q} = \kappa_{r} \bar{\rho} c_{p} \boldsymbol{\nabla} (\bar{T} + T) + \kappa\bar{\rho}\bar{T} \boldsymbol{\nabla} S + \kappa_{0}\bar{\rho}\bar{T} \frac{\partial \bar{S}}{\partial r}\mathbf{r}  $$ 
with $\kappa_{r}$ the molecular radiation diffusion coefficient, $\kappa_{0}$ the effective thermal diffusivity acting only on the spherically symmetric ($l = 0$) entropy and $c_{p}$ the specific heat at constant pressure. A perfect ideal gas equation is used for the mean state and the fluctuation are linearized:
$$ \bar{P} = (\gamma - 1) c_{p} \bar{\rho} \bar{T} / \gamma$$
$$ \rho / \bar{\rho} = P / \bar{P} - T / \bar{T} = P / \gamma \bar{P} - S / c_{p} $$
with $\gamma = 5/3$ the adiabatic exponent.

The anelastic MHD system of equations requires 12 boundary conditions. Magnetic boundary conditions are perfectly conducting at the lower radial boundary and the magnetic field matches to a potential field in the upper boundary:

\hspace{5pt} $ B_{r}|_{r_{in}} = \frac{\partial}{\partial r} \left(\frac{B_{\theta}}{r} \right)|_{r_{in}} = \frac{\partial}{\partial r} \left(\frac{B_{\varphi}}{r} \right) |_{r_{in}} = 0 $ and $ B_{r}|_{r_{out}} = \nabla \Psi \Rightarrow \Delta \Psi = 0 $

We use an impenetrable and stress free at the top and bottom boundaries. 

$$ v_{r} = \frac{\partial}{\partial r}\left( \frac{v_{\theta}}{r} \right) = \frac{\partial}{\partial r}\left( \frac{v_{\phi}}{r} \right) = 0$$
For the mean state of the entropy $(d\bar{S}/dr)|_{r_{out}} = -3.57 \cdot 10^{-9}$ and $(d\bar{S}/dr)|_{r_{in}} = 9.79 \cdot 10^{-4}$cm K$^{-1}$ s$^{-2}$ for M05 model, $(d\bar{S}/dr)|_{r_{out}} = -9.56 \cdot 10^{-9}$ and $(d\bar{S}/dr)|_{r_{in}} = 9.59 \cdot 10^{-3}$cm K$^{-1}$ s$^{-2}$ for M07 model, $(d\bar{S}/dr)|_{r_{out}} = -3.78 \cdot 10^{-8}$ and $(d\bar{S}/dr)|_{r_{in}} = 1.37 \cdot 10^{-2}$cm K$^{-1}$ s$^{-2}$ for M09 model, $(d\bar{S}/dr)|_{r_{out}} = -3.01 \cdot 10^{-7}$ and $(d\bar{S}/dr)|_{r_{in}} = 1.11 \cdot 10^{-2}$cm K$^{-1}$ s$^{-2}$ for M11 model. Keeping the values of $d\bar{S}/dr|_{r_{in},r_{out}}$ fixed at all times in the simulations further implies that the fluctuating $dS/dr$ is set to zero at both bc’s.

The simulation is focused in the bulk convection zone, avoiding regions too close the stellar surface. We include the tachocline in the models, defined as the transition between rigid to differential rotation, leading to region with strong shear \citep{1992AandA...265..106S}. The tachocline plays an important role in the dynamo process of magnetic field generation in solar-like stars as it was reported in simulations performed by several authors \citep{2006ApJ...648L.157B,0004-637X-735-1-46,2014PASJ...66S...2M,2015ApJ...813...95L,2015arXiv150704434G}.

The code uses a realistic background stratification for the profiles of density ($\bar{\rho}$), temperature ($\bar{T}$), $\nu$, $\kappa$ and $\eta$. The background stratification is derived from a one-dimensional solar structure model CESAM \citep{1997AandAS..124..597M,2002ApJ...570..865B}. In Fig. 1 we show an example of the gradient of entropy along the star radius (a), and the temporal and azimuthally averaged radial energy fluxes balance as luminosities (b) for the model with 1.1 solar mass and 3 times the rotation rate of the Sun.

\begin{figure}[h]
\begin{subfigure}{0.5\textwidth}
  \centering
  \includegraphics[width=1.0\linewidth]{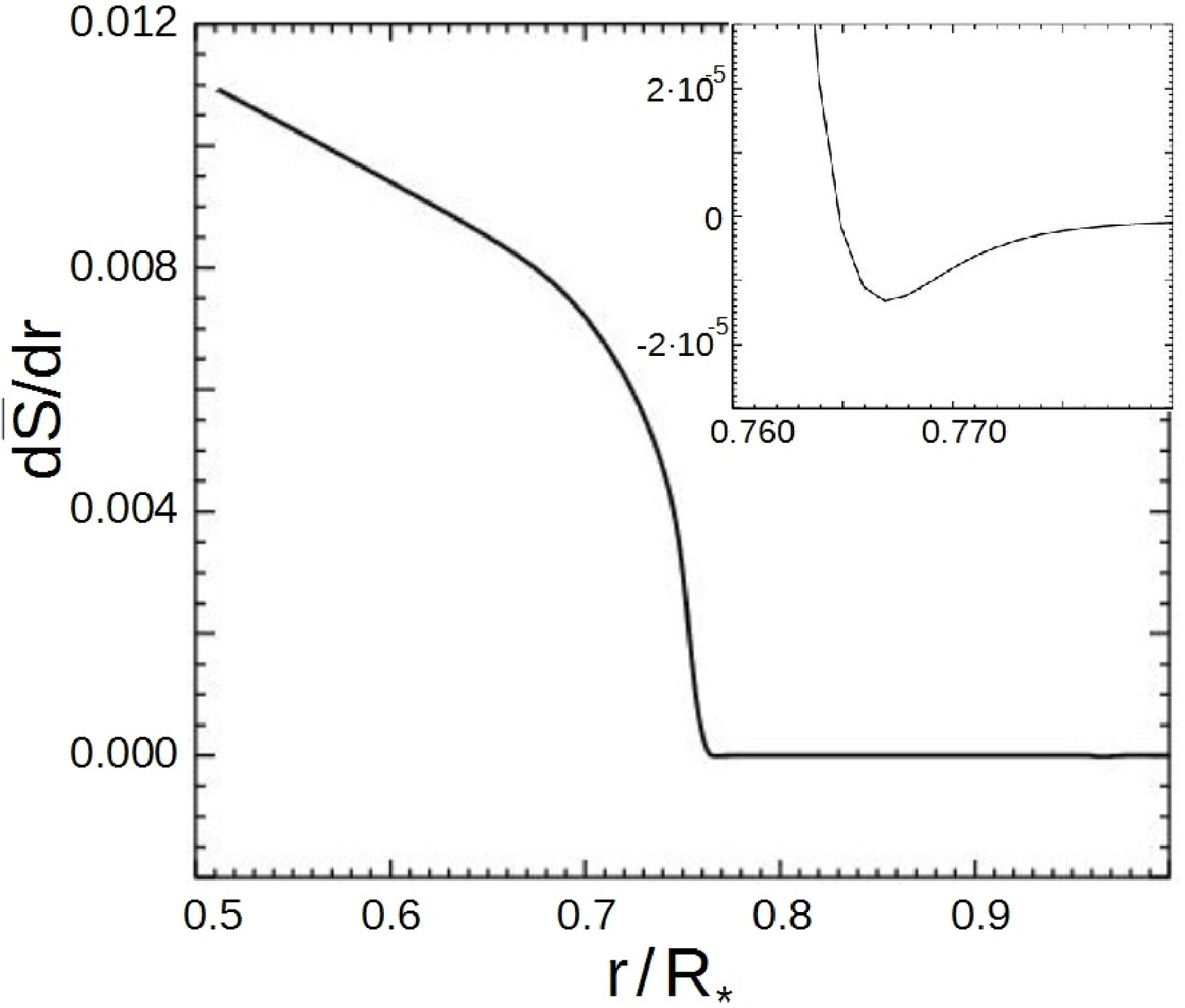}
  \caption{}
  \label{fig1}
\end{subfigure}
\begin{subfigure}{0.502\textwidth}
  \vspace{4pt}
  \includegraphics[width=1.0\linewidth]{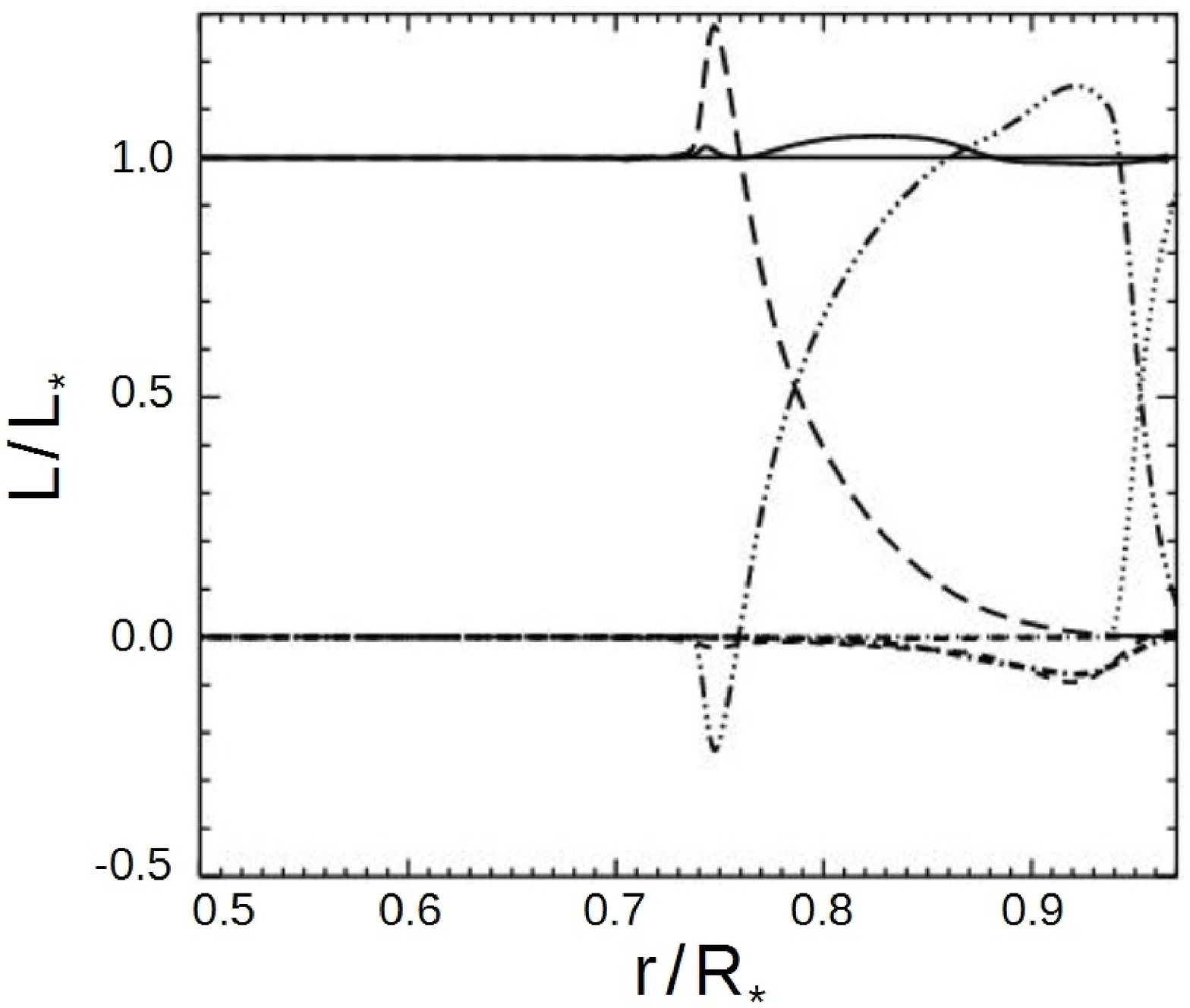}
  \caption{}
  \label{fig2}
\end{subfigure}
\caption{(a) Entropy gradient along the star radius. We expand the region of the tachocline in the top graph. (b) Time and horizontally averaged radial energy fluxes as luminosities (normalized to the star luminosity). The solid line is the total flux, the long dashed line the radiative flux, the dash-dot-dot-dot line the enthalpy flux, the dotted line the conductive entropy flux, the thick dash-doot line the kinetic energy, the dashed line the viscous diffusion flux and the thick dashed line the Poynting flux.}  
\end{figure}

 \begin{table}[h]
 \centering
 \begin{tabular}{c | c c c c c c c}
 Model & $M05_{d1}$ & $M07_{s}$ & $M09_{s}$ & $M09_{d1}$ & $M09_{d3}$ & $M11_{d1}$ & $M11_{d3}$  \\ \hline
 $r_{in}$ ($10^{10}$ cm) & 0.39 & 1.41 & 2.25 & 2.25 & 2.25 & 4.27 & 4.27 \\
 $r_{out}$ ($10^{10}$ cm) & 2.90 & 4.30 & 5.72 & 5.72 & 5.72 & 8.34 & 8.34 \\
 $D$ ($10^{10}$ cm) & 2.51 & 2.89 & 3.47 & 3.47 & 3.47 & 4.07 & 4.07 \\
 $r_{bcr} / r_{out} $ & 0.58 & 0.68 & 0.71 & 0.71 & 0.71 & 0.76 & 0.76 \\
 $\nu$ ($10^{12}$ cm$^2$/s) & 0.45 & 5.31 & 14.3 & 10.01 & 5.81 & 38.01 & 20.95 \\
 $\kappa$ ($10^{13}$ cm$^2$/s) & 0.18 & 2.12 & 5.72 & 4.02 & 2.32 & 15.20 & 8.78 \\
 $\eta$ ($10^{12}$ cm$^2$/s) & 0.22 & 5.31  & 14.30 & 5.03 & 5.81 & 38.01 & 21.95 \\
 $\Omega / \Omega_{\odot}$ & 1 & 0.3 & 0.5 & 1 & 3 & 1 & 3 \\
 $Ta$ ($10^{4}$) & 134.4 & 0.15 & 0.12 & 0.99 & 2.65 & 0.13 & 3.51 \\
 $Ra$ ($10^{6}$) & 19.27 & 0.76 & 0.09 & 0.21 & 1.03 & 0.02 & 0.05 \\
 $Pr$ & 0.25 & 0.25 & 0.25 & 0.25 & 0.25 & 0.25 & 0.25 \\
 $Pr_{m}$ & 2 & 1 & 1 & 2 & 1 & 1 & 1 \\
 $\tau_{\eta}$ (years) & 9.21 & 0.51 & 0.27 & 0.77 & 0.67 & 0.14 & 0.24 \\
 $\tau_{c}$ (years) & 1.01 & 0.45 & 0.30 & 0.30 & 0.30 & 0.16 & 0.16 \\
 $\tilde{v_{r}}$ (m/s) & 8 & 20 & 35 & 35 & 35 & 80 & 80 \\ 
$\tilde{v}$ (m/s) & 66 & 55 & 110 & 154 & 233 & 305 & 484 \\
$\tilde{B}$ ($10^{3}$ G) & 1.1 & 0.8 & 2.3 & 4.2 & 7.2 & 4.5 & 9.5 \\
$Ro $ & 0.31 & 1.84 & 1.68 & 0.84 & 0.28 & 1.17 & 0.39 \\
$Re$ & 370 & 30 & 27 & 53 & 139 & 33 & 90 \\
$Re_{m}$ & 740 & 30 & 27 & 106 & 139 & 33 & 90 \\
$\Lambda$ ($10^{-3}$) & 0.03 & 0.16 & 1.29 & 1.55 & 1.00 & 10.74 & 10.05 \\

\end{tabular}
\caption{Models parameters. $r_{in}$ is the inner radius, $r_{out}$ the outer radius, $r_{bcz}$ the base of the convective zone and $D = r_{out} - r_{in}$ the radial length of the simulation domain. $\nu$, $\kappa$ and $\eta$ are the effective eddy diffusivities of the momentum, heat and magnetic field transport. $\Omega$ is the rotation rate. $Ta = 4\Omega^{2} D^{4}/\nu^{2}$ is the Taylor number. $Ra = (-\partial \bar{\rho}/ \partial S)\Delta Sg D^{3} / \rho \nu \kappa$ is the Rayleigh number with $S$ the entropy and $\rho$ the density ($\bar{\rho}$ means the background state of the density). $Pr = \nu / \kappa$ is the Prandtl number. $Pr_{m} = \nu / \eta$ is the magnetic Prandtl number. $Ro = \tilde{v}/2\Omega D$ is the Rossby number with $\tilde{v}$ the rms velocity in the covection zone. $Re = \tilde{v} D / \nu$ is the Reynolds number with $\tilde{v}$ the rms velocity in the convection zone. $Re_{m} = \tilde{v} D / \eta$ is the magnetic Reynolds number. $\Lambda = \tilde{B}^2/8\pi \bar{\rho} D \tilde{v} \Omega$ is the Elsasser number with $\tilde{B}$ the rms magnetic field strength in the convection zone. $\tau_{\eta} = D^{2} /\pi^2 \eta$ is the ohmic diffusion time. $\tau_{c} = D / \tilde{v_{r}}$ is the overturning convection time with $\tilde{v_{r}}$ the radial component of the averaged velocity in the convection zone. The models resolution is ($N_{r}$, $N_{\theta}$,$N_{\phi}$) 769x256x512 except the model M09 with $3\Omega$ where $N_{\theta}$ is 512.}
\end{table} 

The MHD models are initialized from progenitor hydro models in which a small magnetic field perturbation is introduced (several orders smaller than the final magnetic field observed in the simulation). A first analysis of the progenitor hydro models is done in these references \citet{2011AN....332..897M,2015SSRv..tmp...30B}. The model's resolution is ($N_{r}$, $N_{\theta}$,$N_{\phi}$) 769x256x512 except for model M09 rotating at $\Omega_{*} = 3\Omega_{\odot}$ where $N_{\theta}$ is 512 and $N_{\phi}$ is 1024. In table 1 we indicate the most relevant parameters of the simulations. The 7 models presented in table 1 are named $MA_{xi}$, where $A$ is the mass of the star (in solar masses between $0.5$ to $1.1$) and $i$ the rotation rate of the star (in solar rotation rate). The subindex indicates slow/anti-solar ($x = s$) and solar ($x = d$) differential rotation models (except model $M11_{d1}$ also anti-solar). The density scale heights between the top and the base of the convection zone and  between the top and the bottom of the model are defined as $N_{\rho_{bcz}} = ln (\rho_{out}/\rho_{bcz})$ and $N_{\rho_{tot}} = ln (\rho_{out}/\rho_{in})$. For the $M05$ model $N_{\rho_{bcz}} = 3.25$ and $N_{\rho_{tot}} = 4.70$, $M07$ model $N_{\rho_{bcz}} = 3.48$ and $N_{\rho_{tot}} = 5.78$, $M09$ model $N_{\rho_{bcz}} = 3.31$ and $N_{\rho_{tot}} = 5.99$, $M11$ model  $N_{\rho_{bcz}} = 3.28$ and $N_{\rho_{tot}} = 5.60$.

In the present article we omit the discussion of the dynamo characteristics of the models because this will be the topic of a future communication. We focus the analysis on the differential rotation properties for the various parameters considered.

\section{Large scale flows and energy content in solar-like stars}

In this section we analyze the differential rotation profiles of the models. The aim of the study is to compare the differential rotation profiles of the hydro and MHD models. We show that the presence of magnetic fields leads to different trends of the differential rotation with stellar rotation rate and mass. Following we analyze the kinetic and magnetic energy contained in the models and how they are distributed between their various components.

\subsection{Large scale flows}

We analyze the differential rotation of the simulations that results from the angular momentum redistribution occurring mostly in the convection zone. The panels of figure 2 show an azimuthal plot of the differential rotation averaged over 10 overturning convective times, defined as $\tau_{c} = \int_{r_{bcr}}^{r_{out}} dr / \tilde{v_{r}}$ with $\tilde{v_{r}}$ the radial component of the rms velocity in the convection zone, $r_{out}$ and $r_{bcz}$ the top and the base of the convective layer. We observe that for the simulations $M07_{s}$, $M09_{s}$ and $M11_{d1}$ (the figure of the $M11_{d1}$ model is not shown) there is an anti-solar differential rotation, with the poles rotating faster than the equator, while the other cases show a solar-like differential rotation. There is almost no asymmetry in the profiles between the North and South Hemispheres, as expected when the average is performed over an interval long enough with respect to the convective overturning time. We also display radial cuts of the rotation for the MHD cases (red lines) and Hydro progenitor cases (black lines). In cases rotating 1 and 3 times the solar rotation rate, the poles (latitudes of 75$^{o}$ and higher) are speed up when the magnetic field is present, effect that is stronger as the rotation rate increases. For the $M05_{d1}$ and $M09_{d1}$ models the poles rotates $18$ and $21$ nHz faster in the MHD simulation while for the $M09_{d3}$ and $M11_{d3}$ models the rotation rates increases $155$ and $200$ nHz, pointing out that the effect of the magnetic feedback is larger if the mass of the star increases (larger convective velocity hence larger magnetic field and Maxwell stresses, since $ME \sim KE$). The anti-solar cases show the same behavior, the poles are speed up when the magnetic field is included in the simulation due to the joint action of Maxwell and Reynolds stresses in high latitude regions. If we average the rotation rate between the latitudes 60$^{o}$ and 75$^{o}$ we see that for the $M07_{s}$ and $M09_{s}$ models the rotation rate is enhanced by $10$ and $22$ nHz. The model that shows a larger contrast between the differential rotation and the star rotation is the case $M07_{s}$ followed by the models $M05_{d1}$, $M11_{s}$ and $M09_{s}$. The gradient of the rotation in the radial direction nearby the tachocline is weaker in all the MHD simulations compared with hydro cases. Anti-solar differential rotation models show an increase of the star's rotation near the tachocline for the MHD cases (recalling that $\Omega$ decreases further out in the convection zone) and a drop for the hydro simulation compared with the solid body rotation of the core, indicating how the presence of the magnetic field modifies the rotation gradients (see Fig. 4 for Maxwell stresses). 

\begin{figure}[h]
\begin{subfigure}{0.5\textwidth}
  \centering
  \includegraphics[origin=c,width=1.0\linewidth]{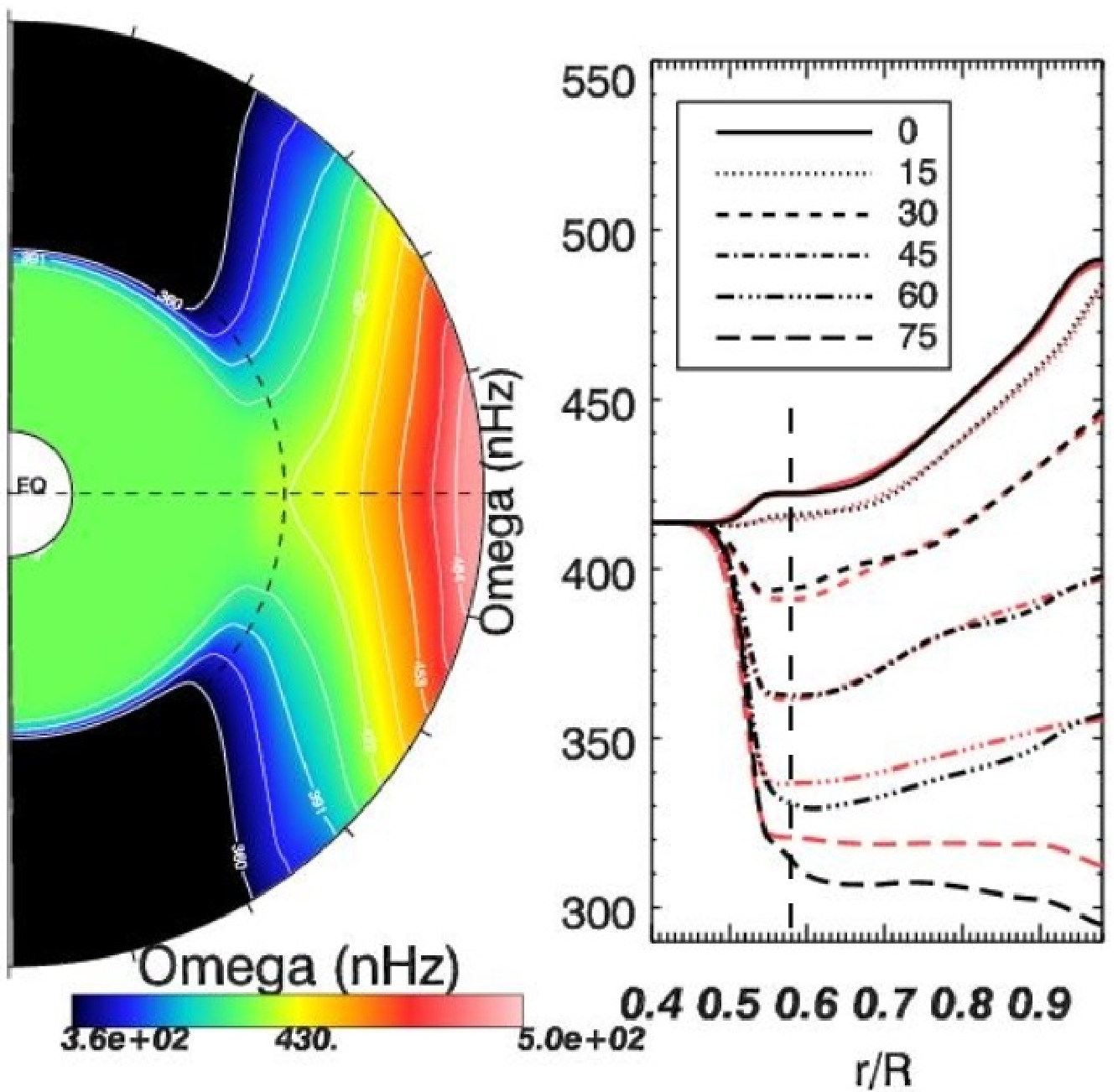}
  \caption{$M05_{d1}$}
  \label{fig1}
\end{subfigure}%
\begin{subfigure}{0.5\textwidth}
  \centering
  \includegraphics[origin=c,width=1.0\linewidth]{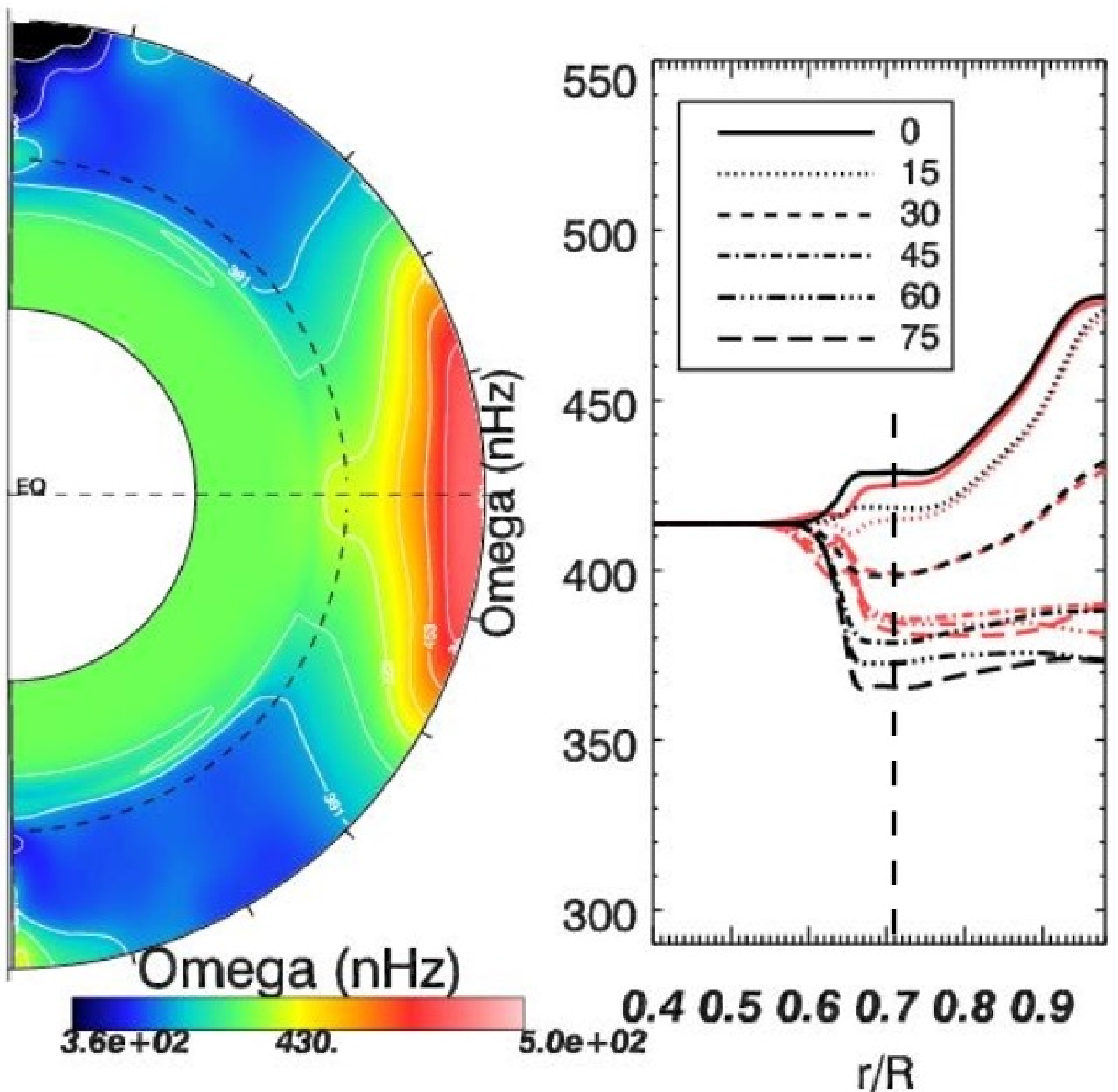}
  \caption{$M09_{d1}$}
  \label{fig2}
\end{subfigure}
\begin{subfigure}{0.5\textwidth}
  \centering
  \includegraphics[origin=c,width=1.0\linewidth]{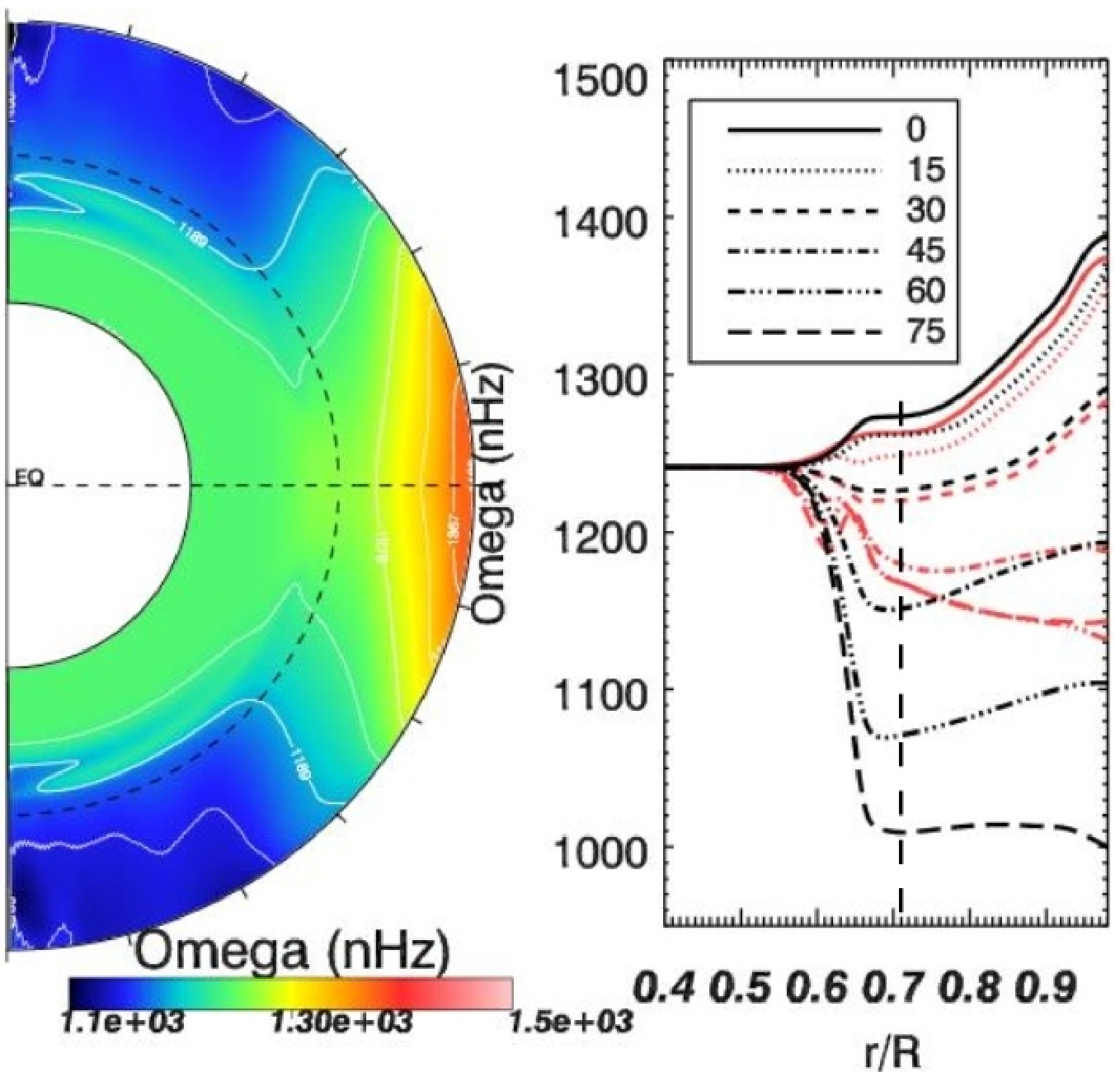}
  \caption{$M09_{d3}$}
  \label{fig3}
\end{subfigure}
\begin{subfigure}{0.5\textwidth}
  \centering
  \includegraphics[origin=c,width=1.0\linewidth]{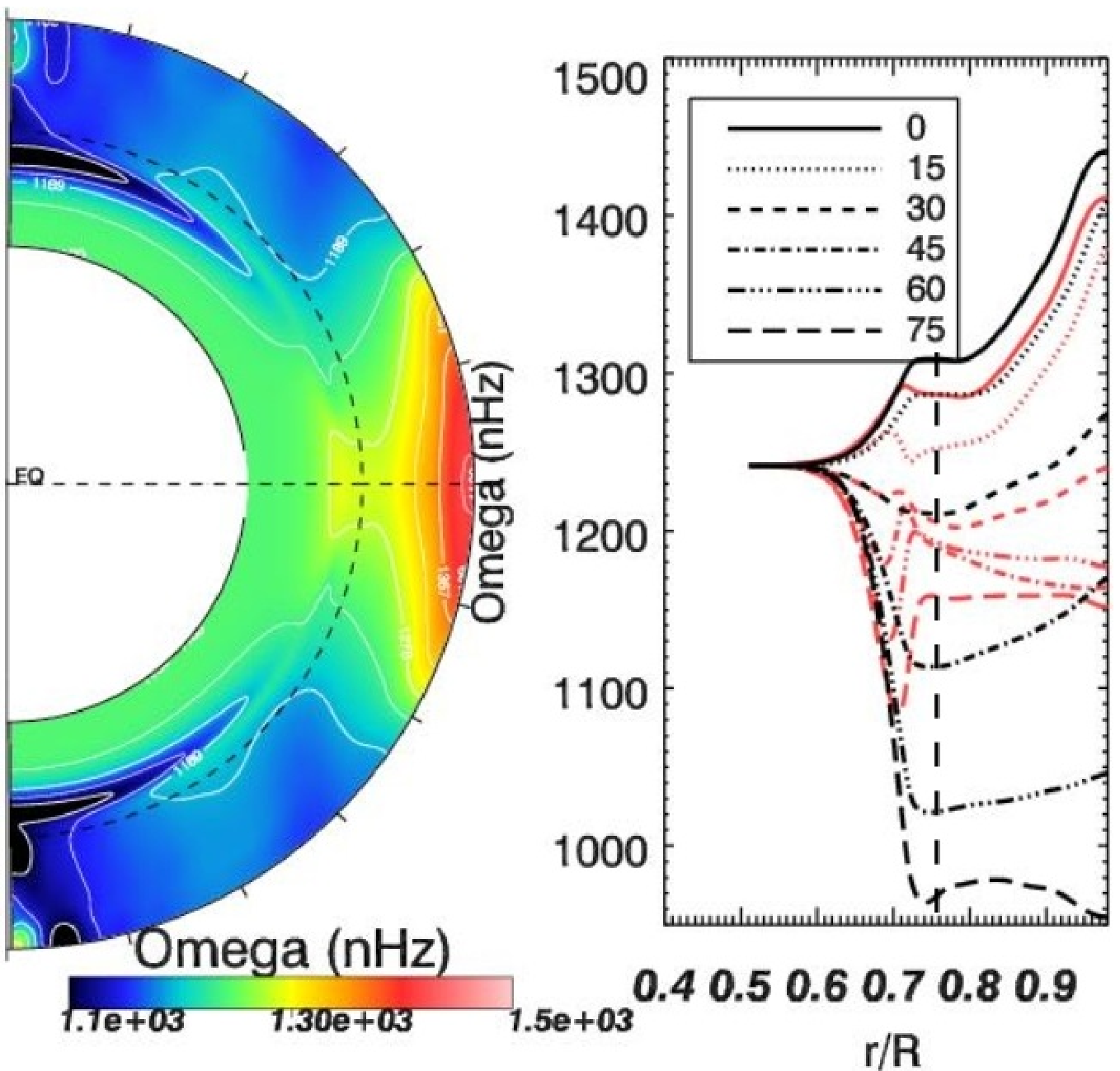}
  \caption{$M11_{d3}$}
  \label{fig4}
\end{subfigure}
\begin{subfigure}{0.5\textwidth}
  \centering
  \includegraphics[origin=c,width=1.0\linewidth]{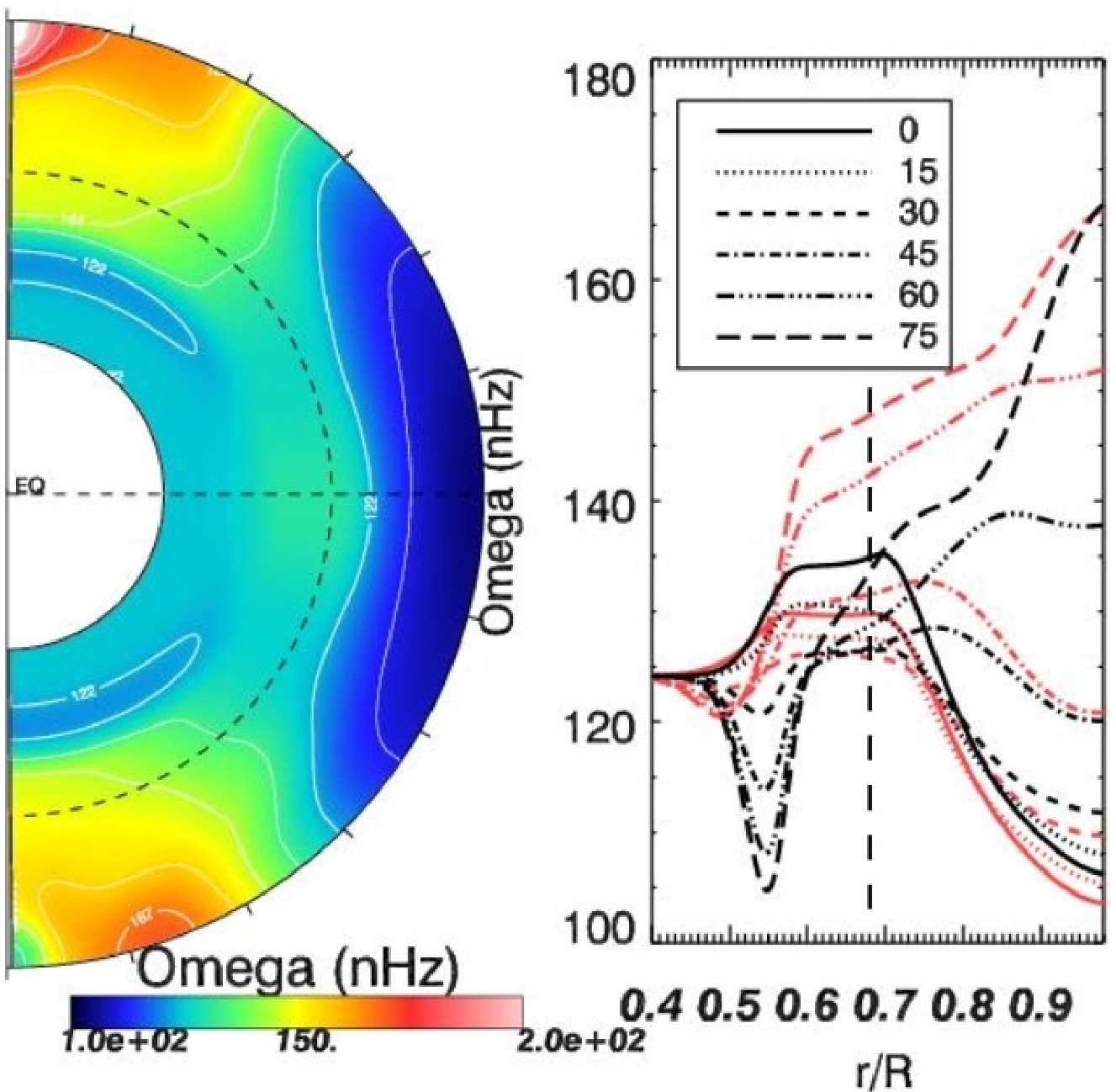}
  \caption{$M07_{s}$}
  \label{fig5}
\end{subfigure}
\begin{subfigure}{0.5\textwidth}
  \centering
  \includegraphics[origin=c,width=1.0\linewidth]{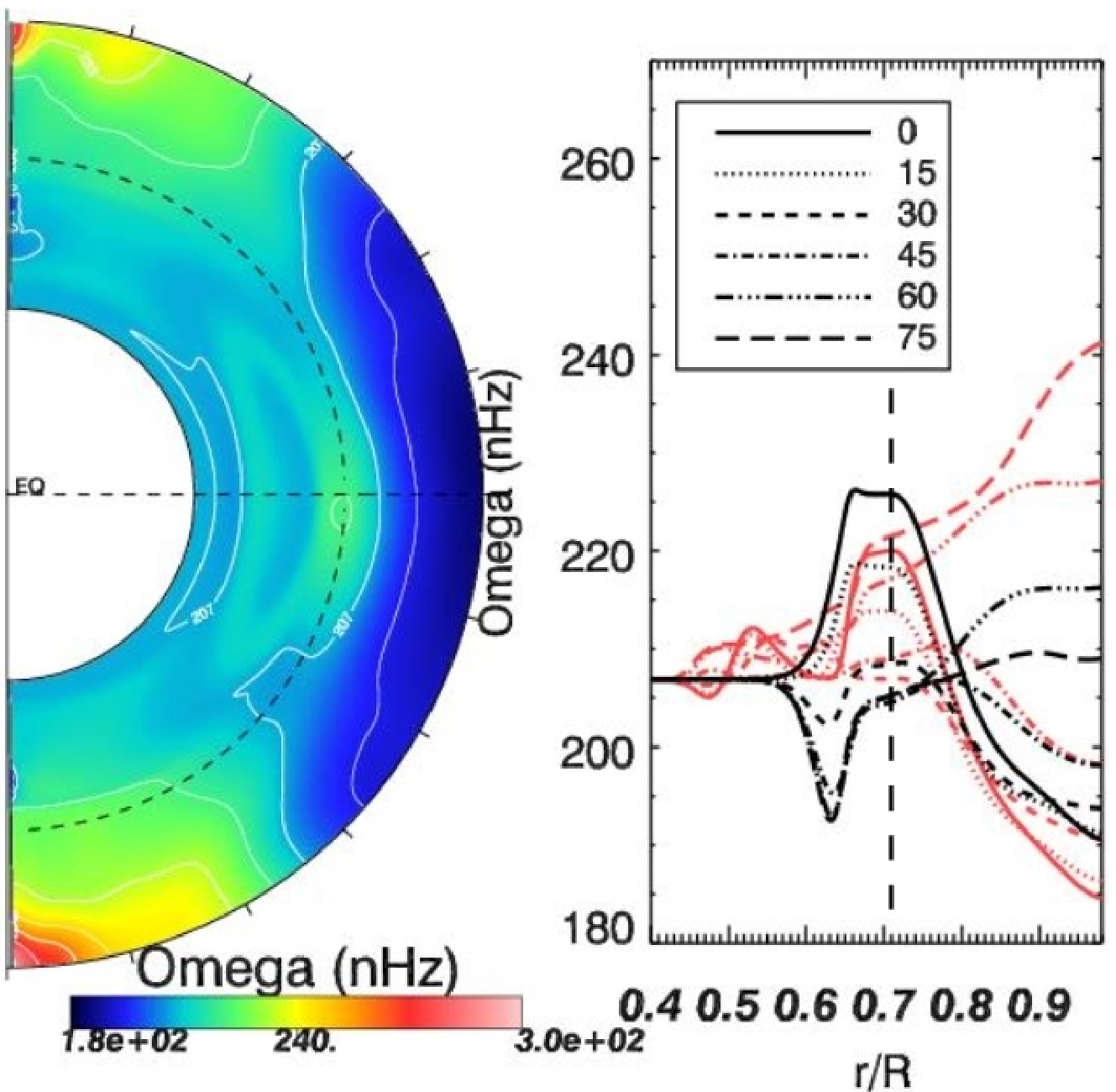}
  \caption{$M09_{s}$}
  \label{fig6}
\end{subfigure}
\caption{Temporal and longitudinal averaged of the angular velocity profiles during 10$\tau_{c}$ and radial cuts from the equator to the latitude 75$^{o}$ each 15$^{o}$ (black lines are the hydro cases and the red lines the MHD cases) between $0.4$ and $0.96$ $r/r_{*}$ ($r/r_{*} = 0.4$ is not necessarily $r = r_{bcr}$). The long dashed vertical line on the right hand side panel of each cases shown the basis of the convective layer.}
\end{figure}

Figure 3 shows the trends of the absolute value of the differential rotation and the differential rotation kinetic energy (DRKE) versus the models Rossby number (graphs A and B). The absolute value of the differential rotation drops when Ro increases, undergoing a transition to the anti-solar differential rotation models if Ro is of the order of 1 or larger with $\Delta \Omega \propto (R_{o})^{-0.40 \pm 0.20}$ (we include the standard error of the fit). The DRKE drops too when Ro increases and it is smaller than $10^{6}$ erg$\cdot$cm$^{-3}$ in the anti-solar models because the models are more dominated by non axisymmetric convection (CKE dominant) with $DRKE \propto (R_{o})^{-1.31 \pm 0.07}$. The trends of the absolute value of the differential rotation versus stellar rotation (graph C), for the regression $\Delta \Omega \propto \Omega^{\alpha}$, is $\Delta \Omega \propto \Omega^{0.44 \pm 0.15}$ in the MHD case and $\Delta \Omega \propto \Omega^{0.89 \pm 0.16}$ in the hydro case. The MHD trend is in better agreement with the observations ($\Delta \Omega \propto \Omega^{0.15}$ \citep{2005MNRAS.357L...1B,2013AandA...560A...4R}). The trends of the absolute value of the differential rotation versus stellar mass (graph D) for the regression $\Delta \Omega \propto (M/M_{\odot})^{\alpha}$, is $\Delta \Omega \propto (M/M_{\odot})^{4.19 \pm 6.86}$ in the MHD case and $\Delta \Omega \propto (M/M_{\odot})^{8.68 \pm 3.91}$ in the hydro case. The inclusion of case $M05_{d1}$ increase significantly the spread of some of the fits and we have chosen to exclude it when necessary to get a better $\chi^2$. We are currently investigating the 0.5 solar mass models at various rotation rates to confirm the behavior of very low mass star models. These results will be reported in a future study, including an updated set of model to reduce the gaps in the results, although the current analysis already shows robust trends for the differential rotation. The MHD trend is in better agreement with the observations ($\Delta \Omega \propto T_{eff}^{8.92}$ ($\Delta \Omega \propto M^{5.6}$) \citep{2005MNRAS.357L...1B,2013AandA...560A...4R} or $\Delta \Omega \propto T_{eff}^{8.6}$ ($\Delta \Omega \propto M^{5.4}$) \citep{2007AN....328.1030C}), confirming that more massive stars have larger differential rotation and predicting more accurately the observed trends.

\begin{figure}[h]
\centering
\includegraphics[width=1.0\textwidth]{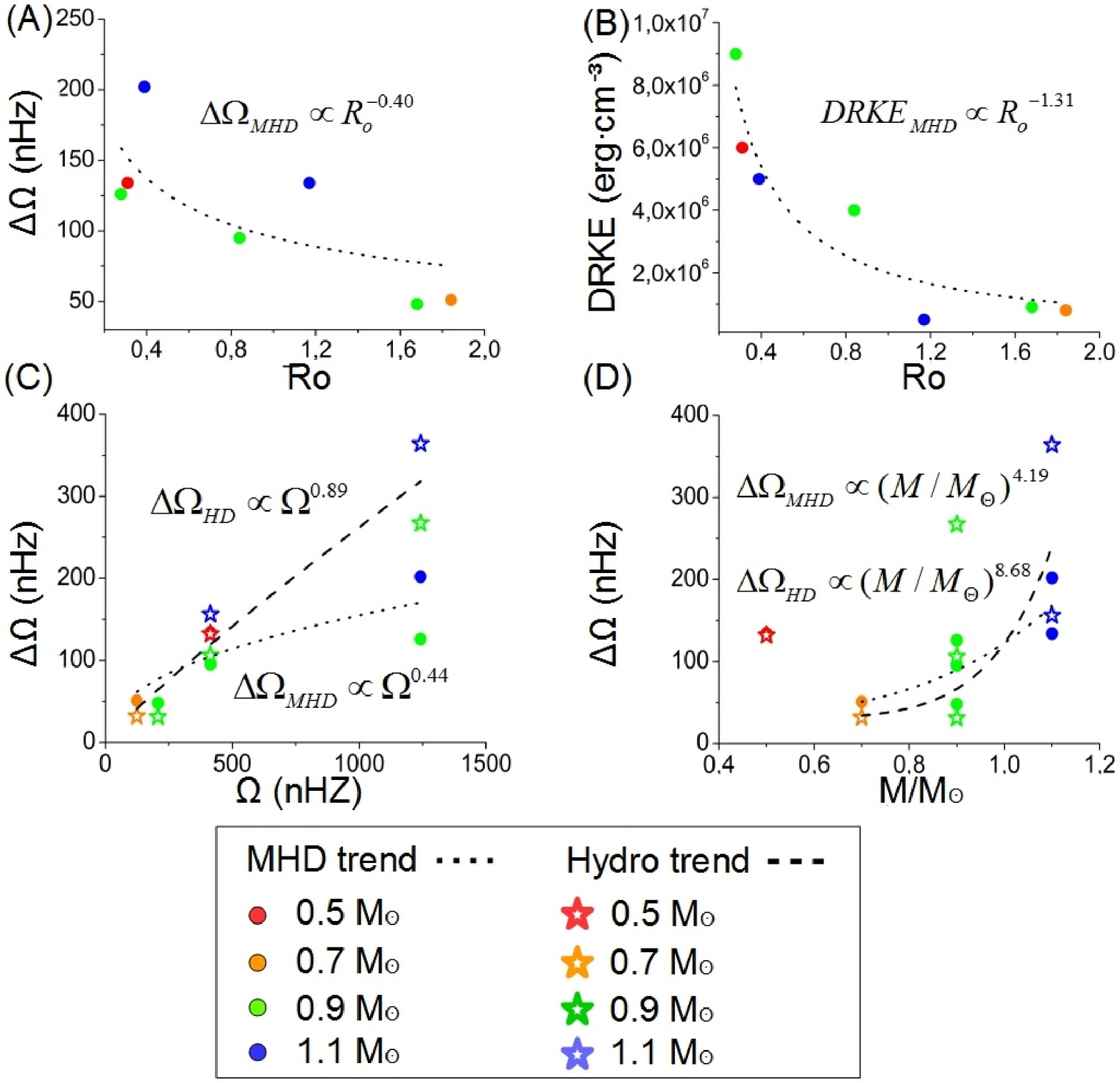}
\caption{Absolute value of the differential rotation between the equator and 60$^{o}$ latitude (A) and differential rotation kinetic energy versus Ro (B). Differential rotation versus rotation (C). Differential rotation versus mass (D). MHD data are the solid circles and hydro data are the empty circles. The dotted (dashed) line shows the linear fit of the MHD (Hydro) data. The data sets of the graphs are fitted to power equations $A = a + \alpha B^{\beta}$ ($\alpha$, $\beta$ and $a$ parameters are fitted).}
\end{figure}

The MHD simulations show how the impact of the magnetic field changes the angular momentum redistribution tending to make the DR more rigid and less sensitive to global parameter changes. In the next section we will perform a detailed analysis of this balance for all the models.

\subsection{Energetic content}

Table 2 indicates the kinetic and magnetic energy of the models averaged in time over the domain (mostly convective zone). Most of the system energy is in form of kinetic energy (KE) for the models $M05_{d1}$, $M07_{s}$ and $M09_{s}$ and the magnetic energy (ME) is at least a $3 \%$ of the total energy in the models $M09_{d1}$, $M09_{d3}$, $M11_{d1}$ and $M11_{d3}$. Between $1$ to $15 \%$ of the star's luminosity is required to maintain $\Omega (r, \theta)$ and the conversion to mean toroidal component of magnetic energy (TME) depends of the Elsasser number of each model. A detailed analysis of the energy transfer including physical explanations is performed in \citet{2011ApJ...742...79B,2015SSRv..196..101B}. The differential rotation kinetic energy component (DRKE) is dominant in the models $M09_{d3}$, $M11_{d3}$ and $M05_{d1}$, while for the models $M07_{s}$, $M09_{s}$ and $M11_{d1}$ it is the turbulent convective kinetic energy (CKE). Only in the model $M09_{d1}$ there is almost the same amount of energy in both components. The meridional circulation kinetic energy (MCKE) is negligible in all cases except for the $M09_{s}$ model where it is a $12 \%$ of the total KE. The mean toroidal component of magnetic energy (TME) is dominant in all simulations, followed by the fluctuation of magnetic energy (FME) that accounts for $20 \%$ of the magnetic energy (value than reaches $30\%$ in $M11_{d1}$ model). The mean poloidal component of the magnetic energy (PME) reaches only $2\%$ of the total energy, except in the anti-solar models $M09_{s}$ ($23\%$) and $M11_{d1}$ ($7\%$). The models with larger differential rotation at the top of the convection zone (calculated as $\Delta \Omega =  \Omega (\theta = 0^{o}) - \Omega (\theta = 60^{o})$) are $M09_{d3}$, $M11_{d1}$ and $M11_{d3}$. The largest ratio between the differential rotation and the stellar rotation is observed for the anti-solar and low mass models. Only the models $M07_{s}$ and $M09_{s}$ show a larger ratio in the MHD models compared with the hydro simulations.

 \begin{sidewaystable}[h]
 \centering
 \begin{tabular}{c | c c c c c c c}
 Model & $M05_{d1}$ & $M07_{s}$ & $M09_{s}$ & $M09_{d1}$ & $M09_{d3}$ & $M11_{d1}$ & $M11_{d3}$  \\ \hline
 KE ($10^{6}$) & 49.8 ($>99\%$) & 5.1 ($>99\%$) & 4.8 ($99\%$) & 6.3 ($97\%$) & 10.0 ($95\%$) & 2.4 ($71\%$) & 7.1 ($92\%$) \\
 DRKE ($10^{6}$) & 42.6 ($86\%$) & 1.0 ($20\%$) & 0.7 ($15\%$) & 3.1 ($49\%$) & 8.9 ($89\%$) & 0.7 ($29\%$) & 5.8 ($82\%$) \\
 CKE ($10^{6}$) & 5.4 ($11\%$) & 4.0 ($78\%$) & 3.5 ($73\%$) & 3.1 ($50\%$) & 1.0 ($10\%$) & 1.7 ($59\%$) & 1.2 ($17\%$) \\
 MCKE ($10^{6}$) & 1.8 ($3\%$) & 0.1 ($2\%$) & 0.6 ($12\%$) & 0.1 ($1\%$) & 0.1 ($1\%$) & 0.0 ($>1\%$) & 0.1 ($1\%$) \\
 ME ($10^{5}$) & 0.14 ($<1\%$) & 0.45 ($<1\%$) & 0.39 ($1\%$) & 2.13 ($3\%$) & 5.11 ($5\%$) & 3.77 ($29\%$) & 6.02 ($8\%$) \\
 PME ($10^{5}$) & 0.00 ($<1\%$) & 0.01 ($2\%$) & 0.09 ($23\%$) & 0.02 ($1\%$) & 0.09 ($2\%$) & 0.28 ($7\%$) & 0.12 ($2\%$) \\
 TME ($10^{5}$) & 0.11 ($79\%$) & 0.35 ($78\%$) & 0.22 ($56\%$) & 1.92 ($90\%$) & 4.06 ($79\%$) & 2.23 ($60\%$) & 5.03 ($84\%$) \\
 FME ($10^{5}$) & 0.03 ($21\%$) & 0.09 ($20\%$) & 0.08 ($21\%$) & 0.19 ($9\%$) & 0.96 ($19\%$) & 1.26 ($33\%$) & 0.83 ($14\%$) \\
$\Delta \Omega$ (nHz) MHD & 134 & -51 & -48 & 95 & 126 & -134 & 202 \\
$\Delta \Omega$ (nHz) HD & 132 & -32 & -31 & 106 & 267 & -156 & 364 \\
$\Delta \Omega / \Omega$ $(\%)$ (MHD) & 32 & 41 & 23 & 23 & 10 & 32 & 16 \\
$\Delta \Omega / \Omega$ $(\%)$ (HD) & 32 & 26 & 15 & 26 & 21 & 38 & 29 \\
\end{tabular}
\caption{Time averaged of the kinetic energy (KE) divided in axisymmetric  differential rotation (DRKE), non axisymmetric  convective (CKE) and axisymmetric  meridional circulation (MCKE) components. The magnetic energy (ME) is divided in toroidal (TME) and poloidal (PME) components. Energy units are ergs cm$^{-3}$. Time averaged differential rotation and normalized values by the star's rotation $\Delta \Omega / \Omega$ for the hydro and MHD simulations.}
\end{sidewaystable}

\section{Angular momentum balance}

Since we wish to focus this paper on stellar DR we now assess which mechanisms maintain it. Hence in this section we study the main physical processes that redistribute the angular momentum in the convective layer. We show the angular momentum balance of the models $M09_{s}$ and $M09_{d3}$ as an example of anti-solar and solar differential rotation cases (see Fig 4). The angular momentum transport can be described by the mean radial $F_{r}$ and latitudinal $F_{\theta}$ angular momentum fluxes \citep{2000ApJ...533..546E,2004ApJ...614.1073B}. We integrate these values in colatitude and in radius, to calculate the net fluxes through respectively cones of various angles and concentric spheres of various radii:

$$ \mathcal{F}_{r}(r) = \int^{\pi}_{0} F_{r}(r,\theta)r^{2}sin\theta d\theta $$
$$ \mathcal{F}_{\theta}(\theta) = \int^{r_{out}}_{r_{in}} F_{\theta}(r,\theta)rsin\theta dr $$
decomposing this transport in viscous diffusion, turbulent Reynolds stresses, meridional circulation (axisymmetric Reynolds stresses from differential rotation + Coriolis), axisymmetric and turbulent Maxwell stresses contributions. Figure 4 shows the $\mathcal{F}_{r}$ and $\mathcal{F}_{\theta}$ integrated angular momentum fluxes for the models $M09_{s}$ and $M09_{d3}$.

\begin{figure}[h]
\begin{subfigure}{0.5\textwidth}
  \centering
  \includegraphics[width=1.0\linewidth]{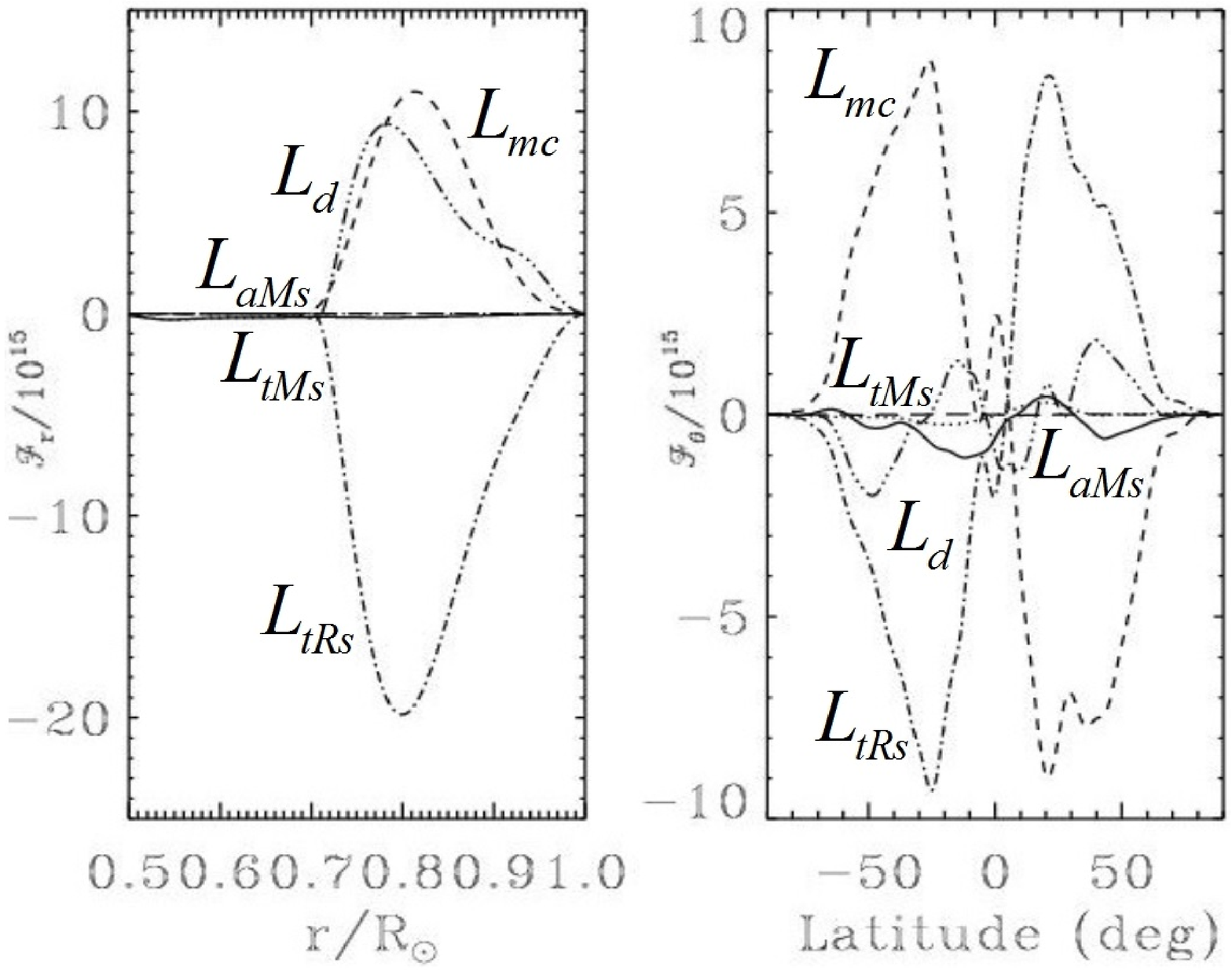}
  \caption{$M09_{s}$ (anti-solar)}
  \label{fig2}
\end{subfigure}
\begin{subfigure}{0.5\textwidth}
  \centering
  \includegraphics[width=1.0\linewidth]{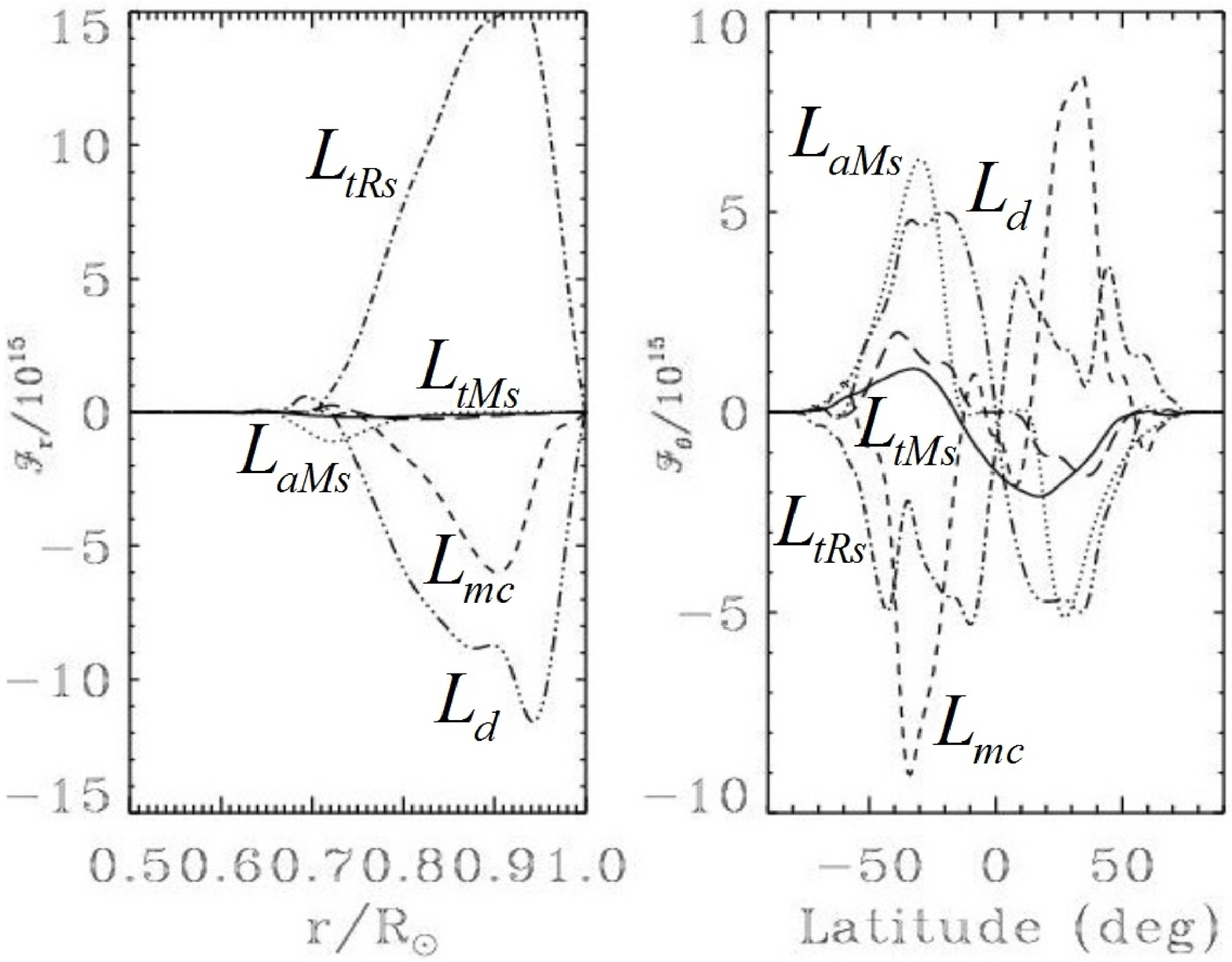}
  \caption{$M09_{d3}$ (solar)}
  \label{fig3}
\end{subfigure}
\caption{Time average of the latitudinal line integral of the angular momentum flux $\mathcal{F}_{r}(r)$ (on the left) and radial line integral of the angular momentum flux $\mathcal{F}_{\theta}(\theta)$ (on the right). Model $M09_{s}$ (top graphs) and $M09_{d3}$ (bottom graphs). The fluxes are decomposed in different contributions: viscous diffusion ($L_{d}$ dash-dot-dot-dot line), turbulent Reynolds stresses ($L_{tRs}$, dash-dot line), meridional circulation ($L_{mc}$, dashed line), axisymmetric ($L_{aMs}$, dotted line) and turbulent ($L_{tMs}$, long dashed line) Maxwell stresses. The solid line shows the total fluxes. The flux is averaged over 10 $\tau_{c}$.}
\end{figure}

The solid line indicates the total net momentum fluxes, almost null for $\mathcal{F}_{r}(r)$ and small for $\mathcal{F}_{\theta}(\theta)$, pointing out that the simulation is in a robust statiscally stationary state. The viscous diffusion term for $\mathcal{F}_{r}(r)$ angular momentum flux is positive in the anti-solar case but negative in the solar case, indicating that in the anti-solar case there is a radially outward transport of angular momentum while in the solar case the transport is inward, consequence of the different radial gradient of the rotation in the models. In the anti-solar case the outward transport is enhanced by the meridional circulation and compensated by the turbulent Reynolds stresses, opposite to what is realized in the solar case. The role of the Maxwell stresses is larger in the solar case than in the anti-solar model, showing a larger peak of the axisymmetric component near the tachocline because the ME in $M09_{d3}$ model is one order larger than in the $M09_{s}$ case), compensating all the other components. For both models there is a small transfer of angular momentum in the stable radiative region nearby the tachocline.

The $\mathcal{F}_{\theta}(\theta)$ angular momentum flux balance shows a more complicated decomposition between components than $\mathcal{F}_{r}(r)$. In the anti-solar case the dominant terms are the turbulent Reynolds stresses and the meridional circulation, which balance each other, with a smaller input of the viscosity (positive near the equator and negative at high latitudes) and the axisymmetric Maxwell stress (enhancing the turbulent Reynolds stresses near the equator) . In the solar case the dominant component is the meridional circulation, enhanced near the equator and at high latitudes by the turbulent Reynolds stresses, and compensated by the viscous diffusion and the axisymmetric and turbulent Maxwell stresses \citep{2014ApJ...789L..19F}. These results point out that in the solar case the role of the Maxwell stresses is important for the angular momentum balance, leading to a configuration with modulated activity, while in this anti-solar case the role of the magnetic fields is smaller and the modulation of the activity weaker, but large enough to increase the differential rotation of the models. If Maxwell stresses are large enough the star's differential rotation decreases due to the quenching effect of the magnetic fields. The decrease of the differential rotation leads to a weaker magnetic field but the Reynolds stresses are opposed to the drop of the differential rotation. The interplay between Maxwell and Reynolds stresses can drive a non stationary evolution of the differential rotation and the magnetic fields of the star that can show chaotic or regular variabilities (see for instance \citet{2005ApJ...629..461B,2011ApJ...731...69B}).

To summarize, we show in Figure 5 the leading components of the angular momentum balance in all the models as an histogram (except $M11_{1d}$ model), where the $\mathcal{F}_{r}(r)$ angular momentum flux is further integrated in radius and $\mathcal{F}_{\theta}(\theta)$ angular momentum flux in latitude (only over the North Hemisphere). The anti-solar cases share same patterns; integrated $\mathcal{F}_{r}(r)$ shows the balance between viscous diffusion and meridional circulation with the turbulent Reynolds stresses, the opposite scenario than in the solar-like differential rotation models. For the integrated $\mathcal{F}_{\theta}(\theta)$ the turbulent Reynolds stresses and the viscous diffusion are balanced by the meridional circulation, while in the solar-like differential rotation models the balance is more complex, especially for the cases with 3$\Omega$ where the Maxwell axisymmetric stresses have an important role. For a larger stellar rotation, the ratio between ME and KE increases leading to a stronger feedback between the fields and the flows. This is easily understood by Table 1 given the larger value of the Elsasser number for the fastly rotating cases which modifies the scaling of the magnetic fields amplitude \citep{2010SSRv..152..565C,2015SSRv..196..101B}.

\begin{figure}[h]
\centering
\includegraphics[width=1.0\textwidth]{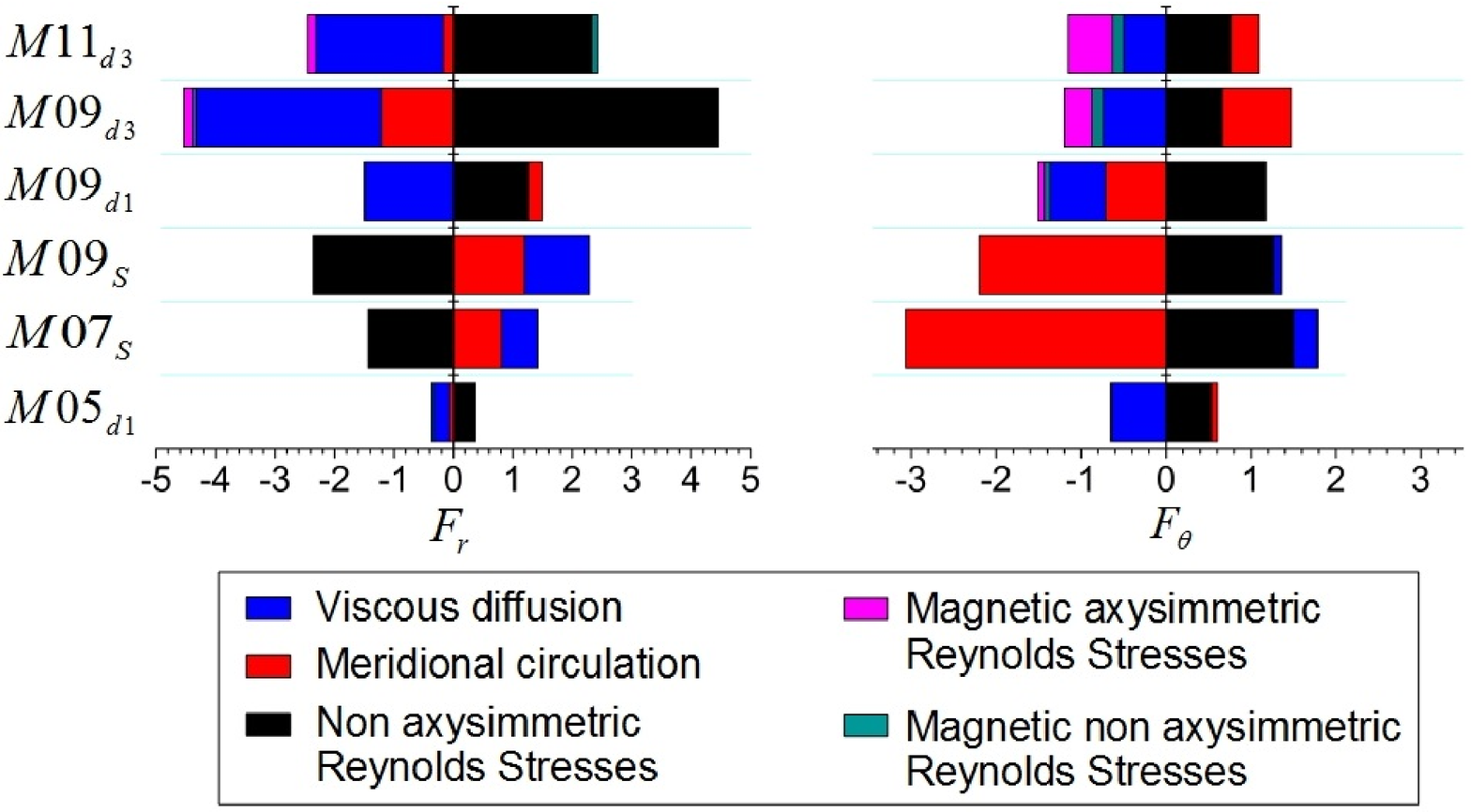}
\caption{Radial (A) and latitudinal (B) net angular momentum fluxes integrated in radius and latitude respectively (normalized to $10^{18}$). The subindex "S" indicates the anti-solar models.}
\end{figure}

Figure 6 shows the temporal and longitudinal average of the meridional circulation for the models $M09_{s}$ and $M09_{d3}$. The meridional circulation is represented by the isocontours of the stream function $\psi$ defined as \citep{2000ApJ...532..593M}:

 $$ \langle \bar{\rho} v_{r} \rangle = \frac{1}{r^{2} sin\theta} \frac{\partial \psi}{\partial \theta} $$
$$ \langle \bar{\rho} v_{\theta} \rangle = - \frac{1}{r sin\theta} \frac{\partial \psi}{\partial r} $$
The meridional circulation is driven by the interplay of the buoyancy and Coriolis forces, pressure gradients, viscosity, Reynolds and Maxwell tensors acting on the zonal flows, leading to deviations in the magnetostrophic equilibrium, consequence of the fluctuating essence of the convection. The system reaction to restore the equilibrium yields a redistribution of the angular momentum by a mechanism called gyroscopic pumping \citep{2006ApJ...641..618M,2007sota.conf..183M,2011ApJ...742...79B}, driving the meridional circulation. In figure 6, the cells with red color indicate counterclockwise rotation and the blue color clockwise rotation. $M09_{s}$ model shows an unicellular meridional circulation, counterclockwise rotating at the North Hemisphere and clockwise at the South Hemisphere. $M09_{d3}$ model shows a multicellular meridional circulation, consequence of a stronger alignment of the cells with the rotation axis as the model rotation rate increases. The Hydro version of $M09_{s}$ and $M09_{d3}$ models shows similar meridional circulation. For the other MHD models, same behavior is observed: anti-solar (retrograde) cases show unicellular flow and prograde ones multi-cellular, increasing the number of cells with the rotation rate \citep{2015ApJ...804...67F}.

\begin{figure}[h]
\centering
\includegraphics[width=0.7\textwidth]{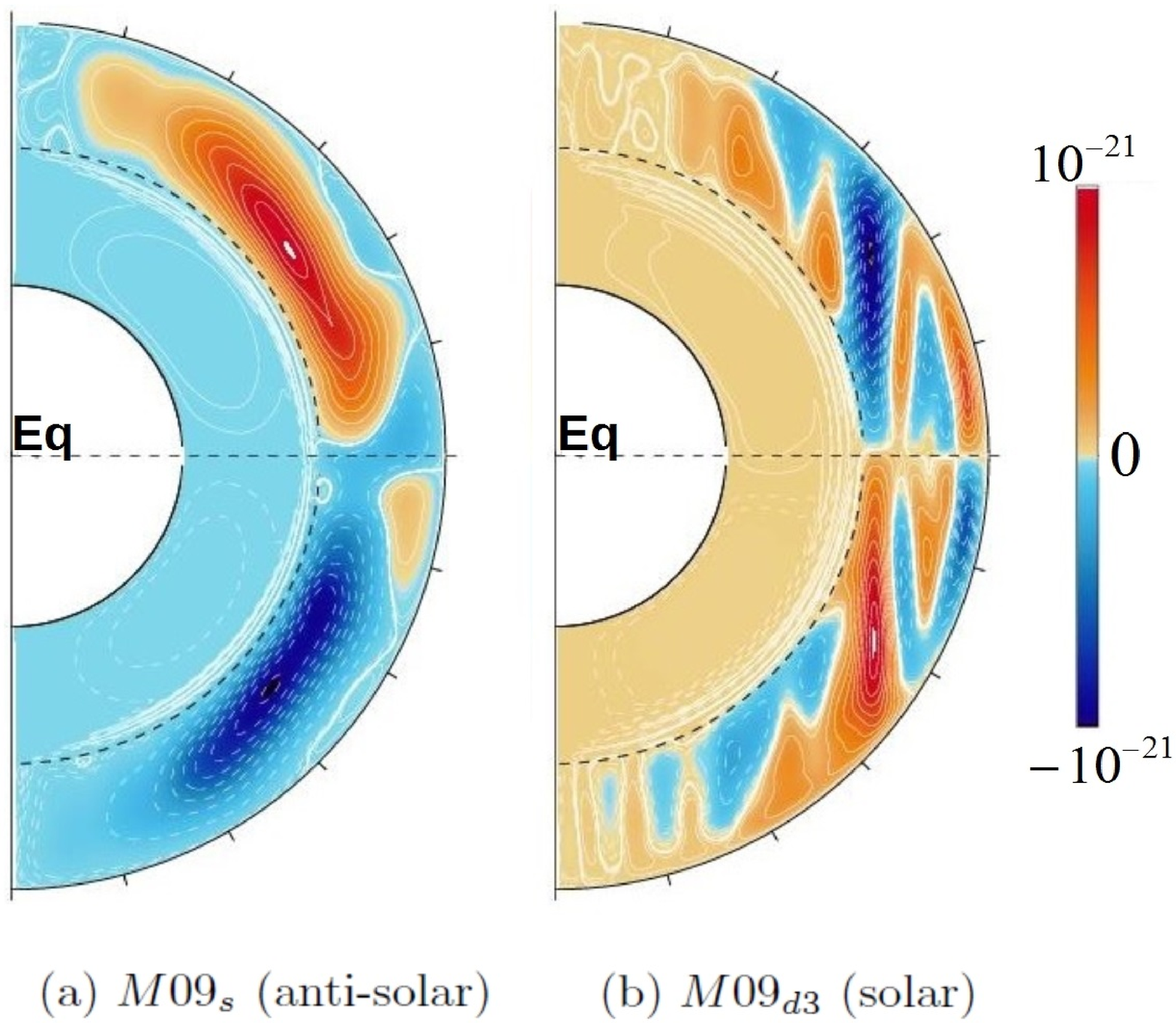}
\caption{Temporal and longitudinal averaged meridional circulation and contours of the stream function $\psi$ for the models $M09_{s}$ and $M09_{d3}$ during 10$\tau_{c}$. Units in g/s.}
\end{figure}

Figure 7 shows an example of time-latitude diagrams of the $\varphi$ component of the magnetic field ($B_{\varphi}$), the torsional oscillations (defined as the difference between the average rotation and the instantaneous rotation of the star in \citet{2000AandA...360L..21C,2003SoPh..213....1S,2005AandA...429..657C,2009SSRv..144..151B}), the temperature fluctuations (defined as the difference between the average temperature and the instantaneous temperature) as well as the turbulent Reynolds stresses and Maxwell stresses components of the latitudinal angular momentum flux at the tachocline ($r/r_{*} = 0.54$, left panels) and at the top of the convective layer ($r = r_{out}$, right panels) for the model $M05_{d1}$. There is a polarity inversion of $B_{\varphi}$ nearby the tachocline between the day 3000 and 4000 of simulated time, while there are successive inversions with much shorter period at the top of the convective layer. The evolution of the $B_{\varphi}$ nearby the tachocline is correlated with the torsional and temperature oscillations at the top of the convection zone, showing that the star's rotation is decelerated and the temperature drops near the equator when $B_{\varphi}$ increases after the polarity inversion. The evolution of the turbulent Reynolds stresses component of the latitudinal angular momentum flux show hints of correlation with the fast-oscillating $B_{\varphi}$ at the top of the convection zone. A longer time-series would though be needed to further confirm this correlation is significant compared to epochs where the oscillating $B_{\varphi}$ is weaker. At the tachocline, the turbulent Reynolds stresses show stable patterns nearby the equator, isolated among hemispheres. The Maxwell stresses components of the latitudinal angular momentum flux at the tachocline follow $B_{\varphi}$ evolution, showing a minimum during the polarity inversion. Near the upper boundary, Maxwell stresses increase with $B_{\varphi}$, indicating a potentially significant feedback effect over the star's differential rotation.

\begin{figure}[h]
\centering
\includegraphics[width=1.0\textwidth]{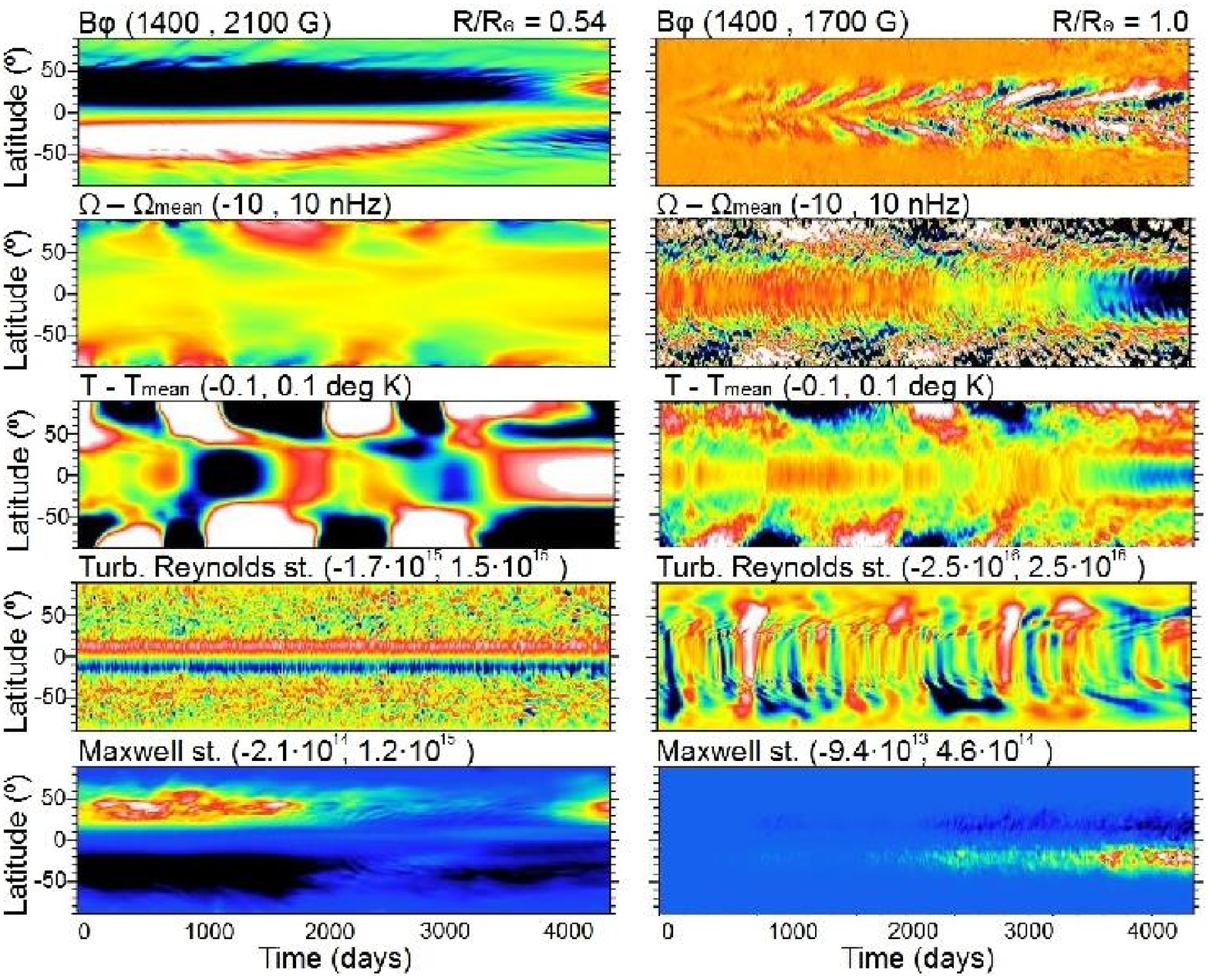}
\caption{Model $M05_{d1}$. Time-latitude diagrams of the $B_{\varphi}$ component of the magnetic field, the torsional oscillations, temperature oscillations as well as the turbulent Reynolds stresses and the Maxwell stresses components of the latitudinal angular momentum flux (dyn$\cdot$cm) at the tachocline ($r/r_{*} = 0.54$, left panels) and the top of the convective layer ($r = r_{out}$, right panels)}.
\end{figure}

\section{Baroclinity and Thermal Wind balance}

The effect of rotation upon convection leads to latitudinal heat transport that establishes gradients in temperature and entropy. To analyze this effect, Figure 8 shows azimuthal plots of temperature and entropy averaged over 10 overturning convective times for the models $M09_{s}$ and $M09_{d3}$.

\begin{figure}[h]
\begin{subfigure}{0.5\textwidth}
  \centering
  \includegraphics[origin=c,width=1.0\linewidth]{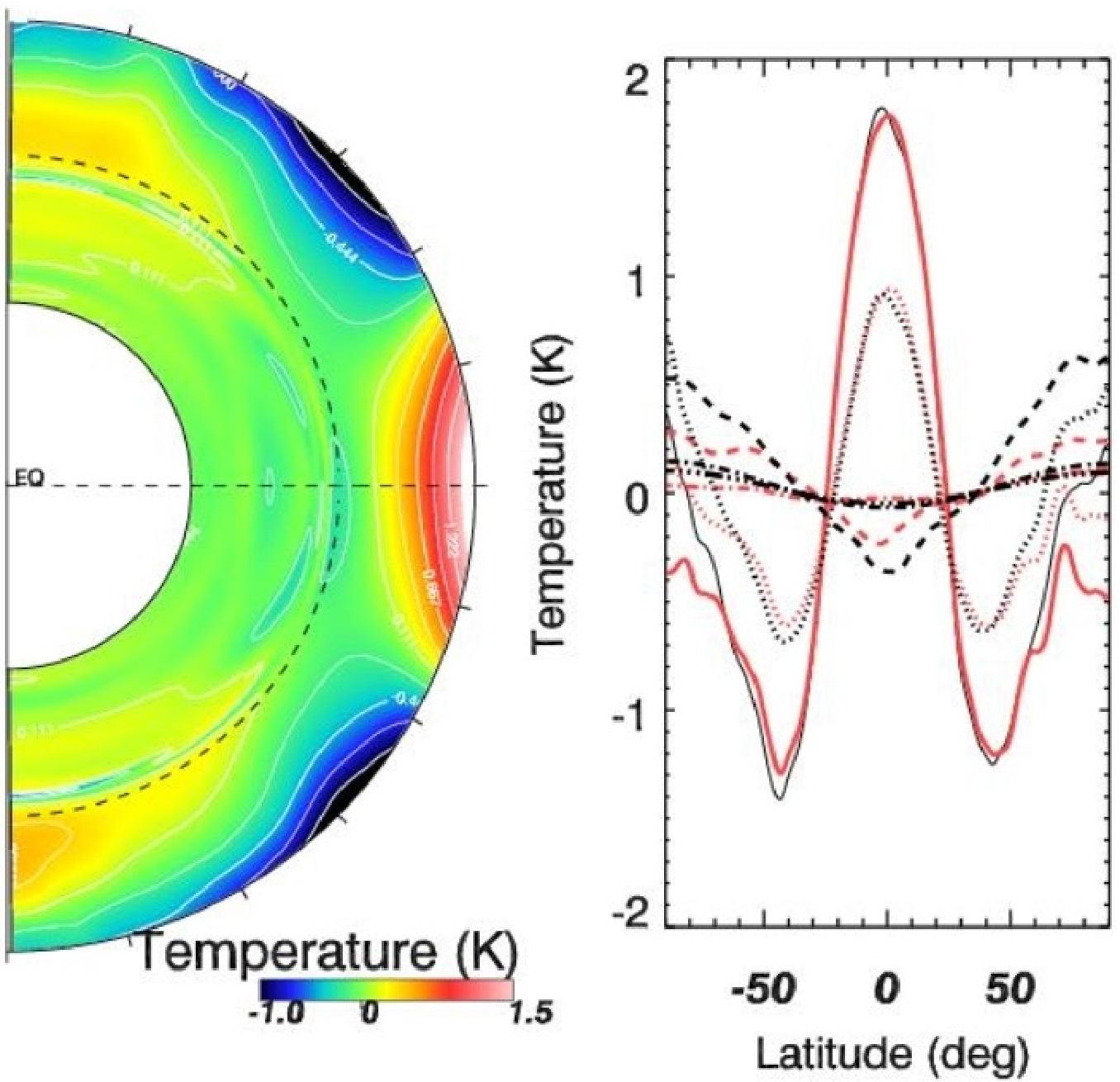}
  \caption{Temperature $M09_{s}$ model}
  \label{fig1}
\end{subfigure}%
\begin{subfigure}{0.5\textwidth}
  \centering
  \includegraphics[origin=c,width=1.0\linewidth]{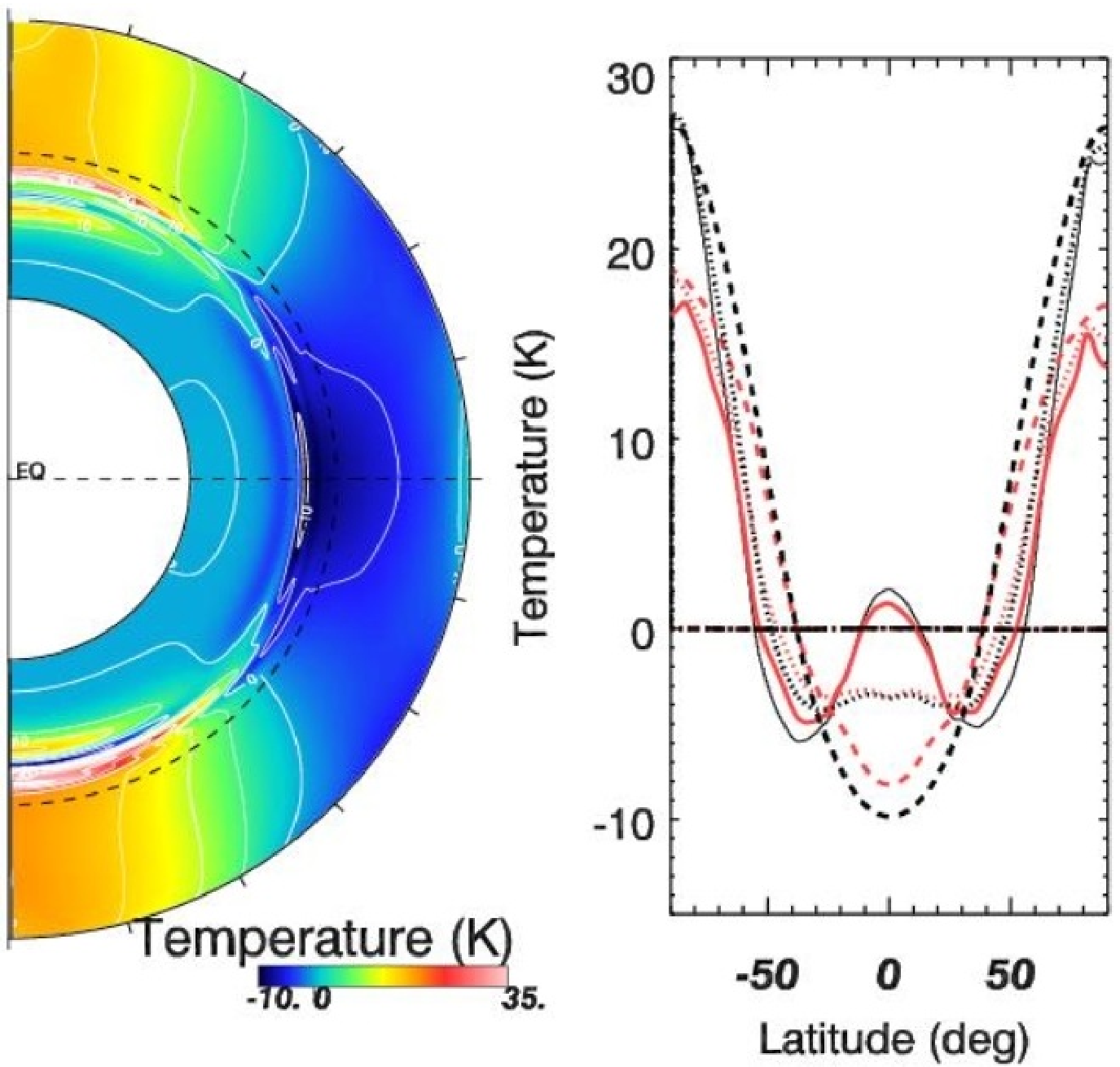}
  \caption{Temperature $M09_{d3}$ model}
  \label{fig2}
\end{subfigure}
\begin{subfigure}{0.5\textwidth}
  \centering
  \includegraphics[origin=c,width=1.0\linewidth]{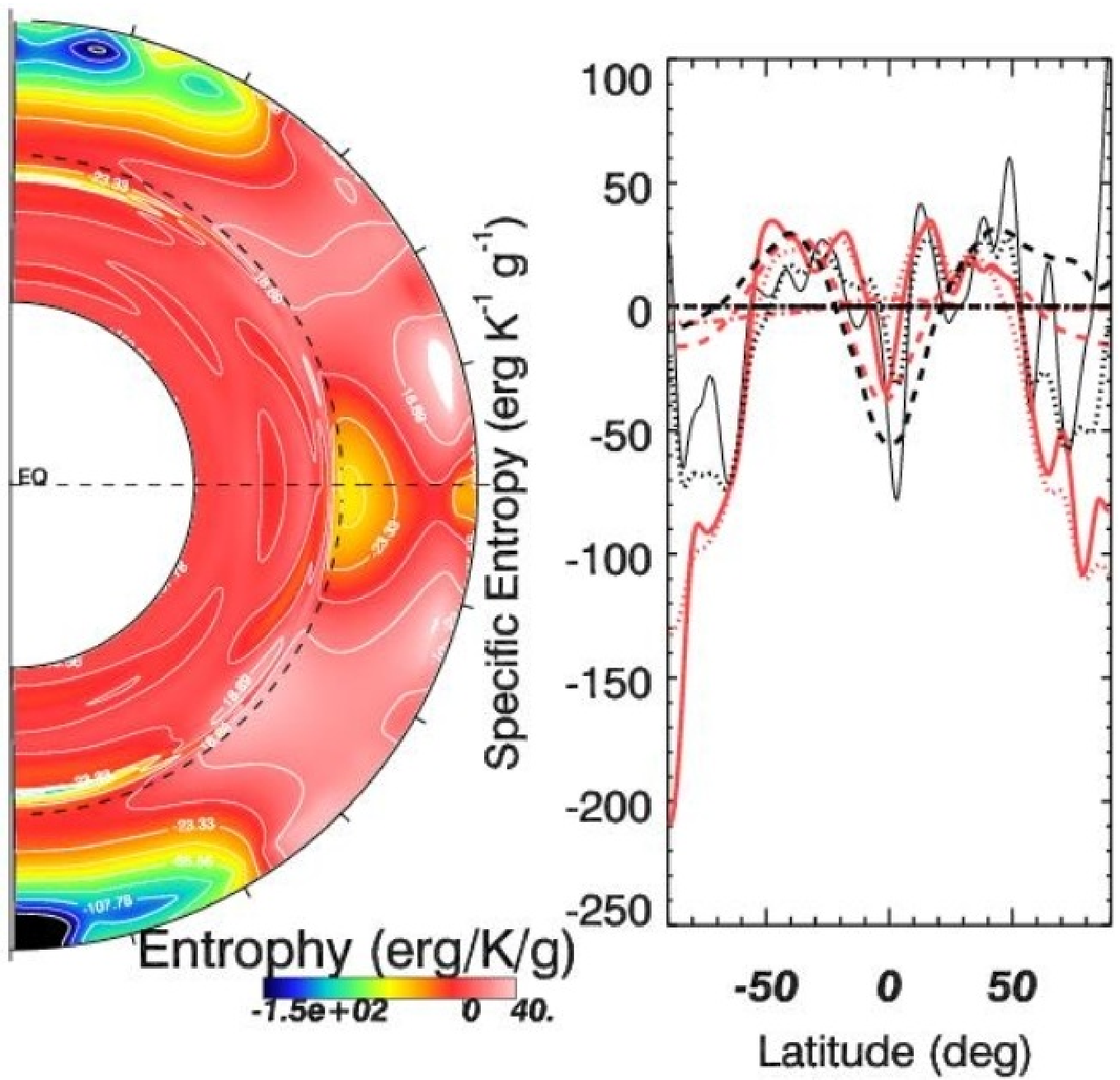}
  \caption{Entropy $M09_{s}$ model}
  \label{fig3}
\end{subfigure}
\begin{subfigure}{0.5\textwidth}
  \centering
  \includegraphics[origin=c,width=1.0\linewidth]{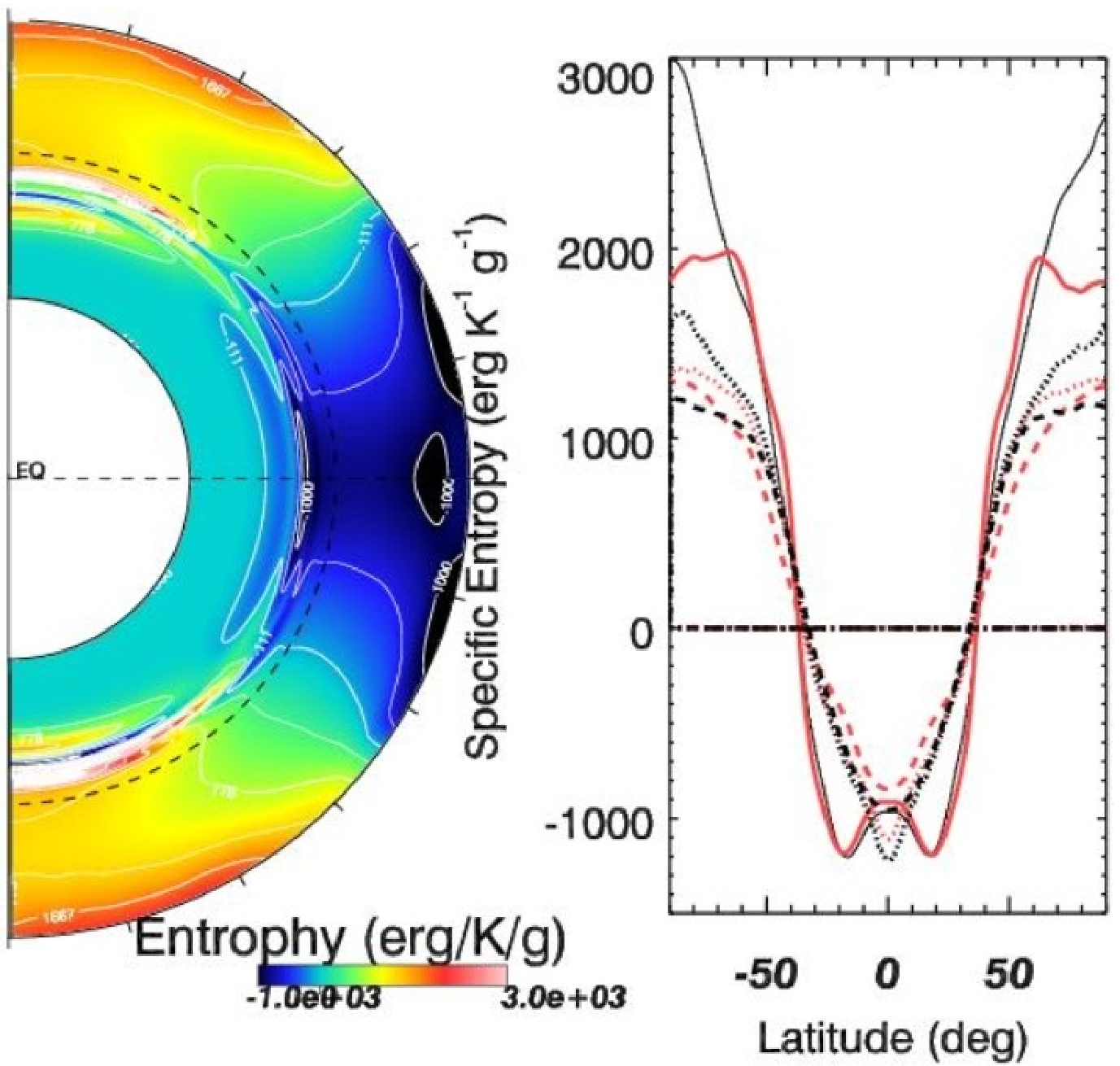}
  \caption{Entropy $M09_{d3}$ model}
  \label{fig4}
\end{subfigure}
\caption{Temporal and longitudinal averaged of temperature and entropy profiles during 10$\tau_{c}$, including latitudinal cuts at radius $r_{out}$ (solid line), $3 \cdot r_{out}/4$ (dotted line), $r_{out}/2$ (dashed line), $r_{out}/4$ (dash-dot line), and $r_{in}$ (dash-dot-dot-dot line) (black lines are the hydro cases and the red lines the MHD cases).}
\end{figure}

There are gradients of entropy and temperature in the convective layer, particularly large in latitude when comparing values at the equator and poles although gradients in radius are large too at high latitudes. The rotation is slower in $M09_{s}$ model reason why the gradients are small compared with $M09_{d3}$ model (because $\frac{\partial \langle S \rangle}{\partial \theta} = \frac{2\Omega_{0} rc_{p}}{g} \frac{\partial \langle v_{\phi} \rangle}{\partial z}$, see \citet{2008ApJ...689.1354B}). Hydro simulations show stronger gradients than the MHD cases in agreement with the overall large angular velocity contrast. Figure 9 shows the temperature and entropy constrast (defined between the latitude $60^{o}$ and the equator) with the star's rotation. The trends for the temperature contrast, defined as $\Delta T \propto (\Omega/\Omega_{\odot})^{\alpha}$, is $\Delta T \propto (\Omega/\Omega_{\odot})^{0.59 \pm 0.25}$ in the MHD case and $\Delta T \propto (\Omega/\Omega_{\odot})^{0.66 \pm 0.26}$ in the hydro case. The trends for the entropy contrast (weighted  by the star's luminosity) is $\Delta S (L_{\odot}/L) \propto (\Omega/\Omega_{\odot})^{0.81 \pm 0.52}$ in the MHD case and $\Delta S (L_{\odot}/L) \propto (\Omega/\Omega_{\odot})^{0.83 \pm 0.81}$ in the hydro case. So we see that the temperature and entropy contrasts increase with the star's rotation in agreement with thermal wind-like balance, but with a weaker trend in the MHD cases.

\begin{figure}[h]
\centering
\includegraphics[width=1.0\textwidth]{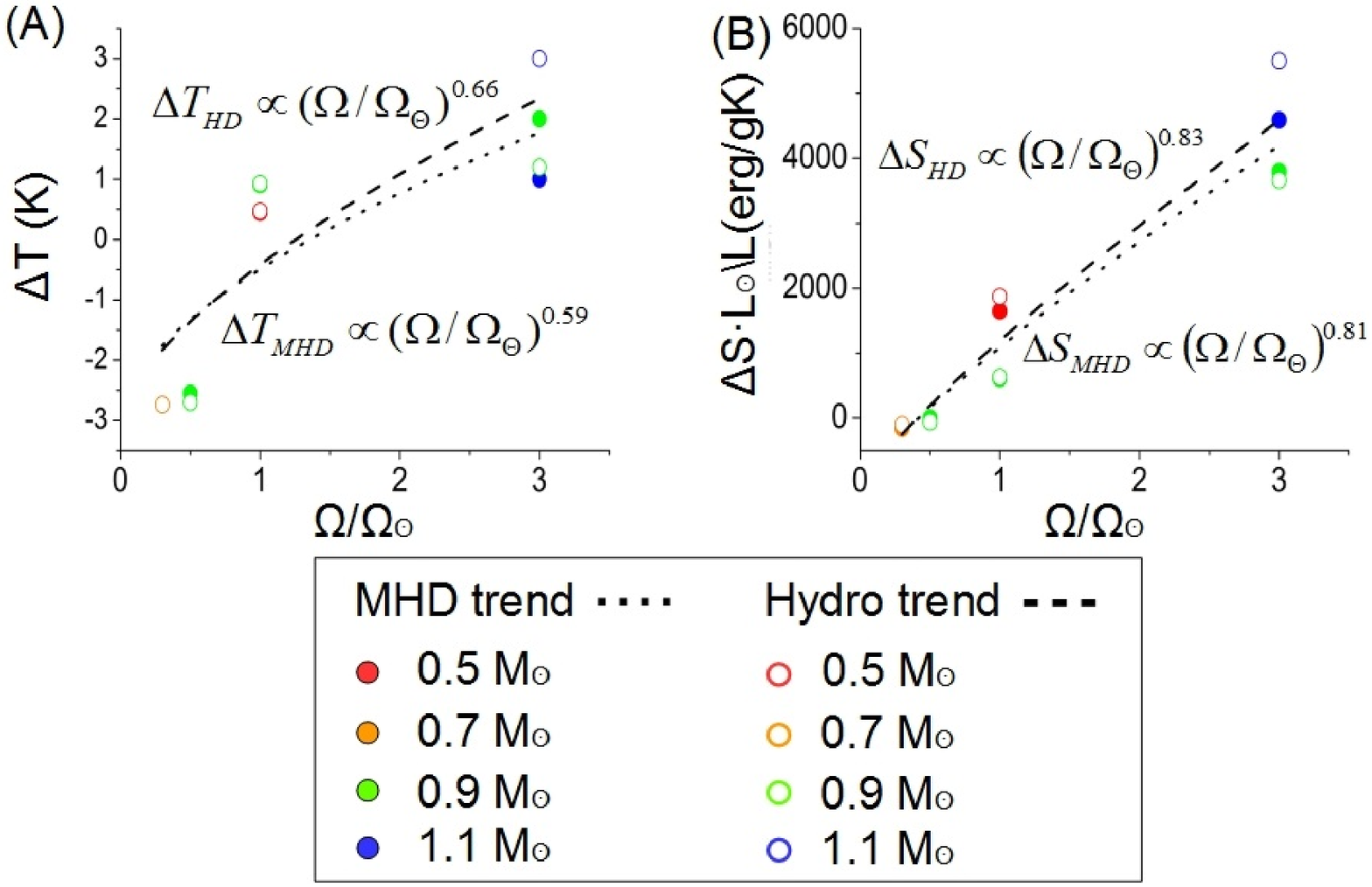}
\caption{Trends of the temperature (A) and entropy contrasts respect to the star's rotation at the top of the convective layer. Entropy contrast is weighted  by the star's luminosity. The data sets of the graphs are fitted to power equations $A = ctte + \alpha B^{\beta}$ ($\alpha$, $\beta$ and ctte parameters are fitted)}
\end{figure}

The presence of gradients in temperature and entropy leads to a mismatch of the iso-surfaces of mean density and pressure that is named baroclinicity and appears in the vorticity equation \citep{1992AandA...265..115Z,2006ApJ...641..618M,2010AandA...510A..33B}. The baroclinic term contributes to breaking Taylor constraint of a cylindrical mean flow yielding more complex (conical) angular velocity profiles \citep{1995AandA...299..446K}. In figures 10 and 11 we show for the models $M09_{s}$ and $M09_{d3}$ if the departure from cylindricity of the differential rotation is accounted for mainly by the baroclinic term or if we must consider other effects. The balance of the mean zonal components of the curl of the momentum (time and azimuthal averaged defined with the symbol $\langle  \rangle$) can be expressed as \citep{2000ApJ...533..546E,2004ApJ...614.1073B,2011AandA...532A..34S}:

$$ 2\Omega_{0}\frac{\partial v_{\phi}}{\partial z} = -\left\langle (\boldsymbol{\omega}\cdot\boldsymbol{\nabla})v_{\phi} + \frac{\omega_{\phi}v_{r}}{r} + \frac{\omega_{\phi}v_{
\theta}cot\theta}{r} \right\rangle + \left\langle (\mathbf{v}\cdot\boldsymbol{\nabla})\omega_{\phi} + \frac{v_{\phi}\omega_{r}}{r} + \frac{v_{\phi}\omega_{\theta}cot\theta}{r} \right\rangle  $$
$$ - \left\langle \omega_{\phi} v_{r} \right\rangle \frac{d ln \bar{\rho}}{dr} + \frac{g}{rc_{p}}\frac{\partial \langle S \rangle}{\partial \theta} + \frac{1}{r\bar{\rho}c_{p}}\frac{d\bar{S}}{dr}\frac{\partial \langle P \rangle}{\partial \theta} + \frac{1}{r} \left[\frac{\partial}{\partial r} (r \langle A_{\theta} \rangle) - \frac{\partial}{\partial \theta} \langle A_{r} \rangle \right] $$
$$ + \left\langle -\frac{1}{c\bar{\rho}}(\mathbf{B}\cdot\boldsymbol{\nabla}) j_{\phi} - \frac{j_{r} B_{\phi}}{c\bar{\rho}r} - \frac{j_{\theta} B_{\phi} cot\theta}{c\bar{\rho}r}                           \right\rangle + \left\langle \frac{1}{c\bar{\rho}}(\mathbf{j}\cdot\boldsymbol{\nabla}) B_{\phi} + \frac{j_{\phi} B_{r}}{c\bar{\rho}r} + \frac{j_{\phi} B_{\theta} cot\theta}{c\bar{\rho}r} \right\rangle $$ 
$$ + \left( \frac{\left\langle B_{r} j_{\phi} \right\rangle}{c \bar{\rho}} - \frac{\left\langle B_{\phi} j_{r} \right\rangle}{c \bar{\rho}} \right) \frac{d ln \bar{\rho}}{dr}      $$
where:
$$\frac{\partial}{\partial z} = cos\theta \frac{\partial}{\partial r} - sin\theta \frac{\partial}{\partial \theta}$$ 
$$A_{r} = \frac{1}{\bar{\rho}} \left[ \frac{1}{r^2}\frac{\partial (r^{2} D_{rr})}{\partial r} + \frac{1}{r sin\theta}\frac{sin\theta D_{\theta r}}{\partial \theta} - \frac{D_{\theta\theta} + D_{\phi\phi}}{r} \right] $$
$$A_{\theta} = \frac{1}{\bar{\rho}} \left[ \frac{1}{r^2}\frac{\partial (r^{2} D_{r\theta})}{\partial r} + \frac{1}{r sin\theta}\frac{sin\theta D_{\theta \theta}}{\partial \theta} + \frac{D_{\theta r} -cot\theta D_{\phi\phi}}{r} \right] $$
with the definitions: $\boldsymbol{\omega} = \boldsymbol{\nabla} \times \mathbf{v}$ the vorticity, $\mathbf{j}$ the current density, $c_{p}$ the specific heat at constant pressure and $D_{ij}$ the viscous stress tensor. The first term in the equation is the stretching by velocity gradients, the second term the advection by the flow, the third term the compressibility, the fourth and fifth terms the baroclinic terms due to non-alignment of density and pressure gradients and of the departures from the adiabatic stratification, the term number 6 the diffusion by viscous stresses and the last 3 terms are the magnetic contributions to the "shear" and "transport" of the magnetic field by the current and compressibility.

We expect that the thermal wind is stronger as the star's rotation increases according to the increase of temperature and entropy gradients. For the anti-solar case the baroclinic term is the leading component in all the domain except near the upper boundary at high latitudes, where the viscous and Reynolds stresses (particularly the advection) are in opposition to the baroclinic term. For the solar case the baroclinic term is more dominant than in the anti-solar model. Only very close to the top of the convective layer there are meaningful signatures of the viscous stresses and the stretching component of the Reynolds stresses, in opposition to the baroclinic term near the equator, as well as the advection component of the Reynolds stresses at mid and high latitudes. The magnetic contributions are mainly negligible with only weak signatures near the tachocline. In summary, for the anti-solar model the Reynolds stresses and viscous effects should be considered especially at high latitudes, while for the solar case the baroclinic term dominates in the entire domain except very close to the surface. 

\begin{figure}[h]
\centering
\includegraphics[width=1.0\textwidth]{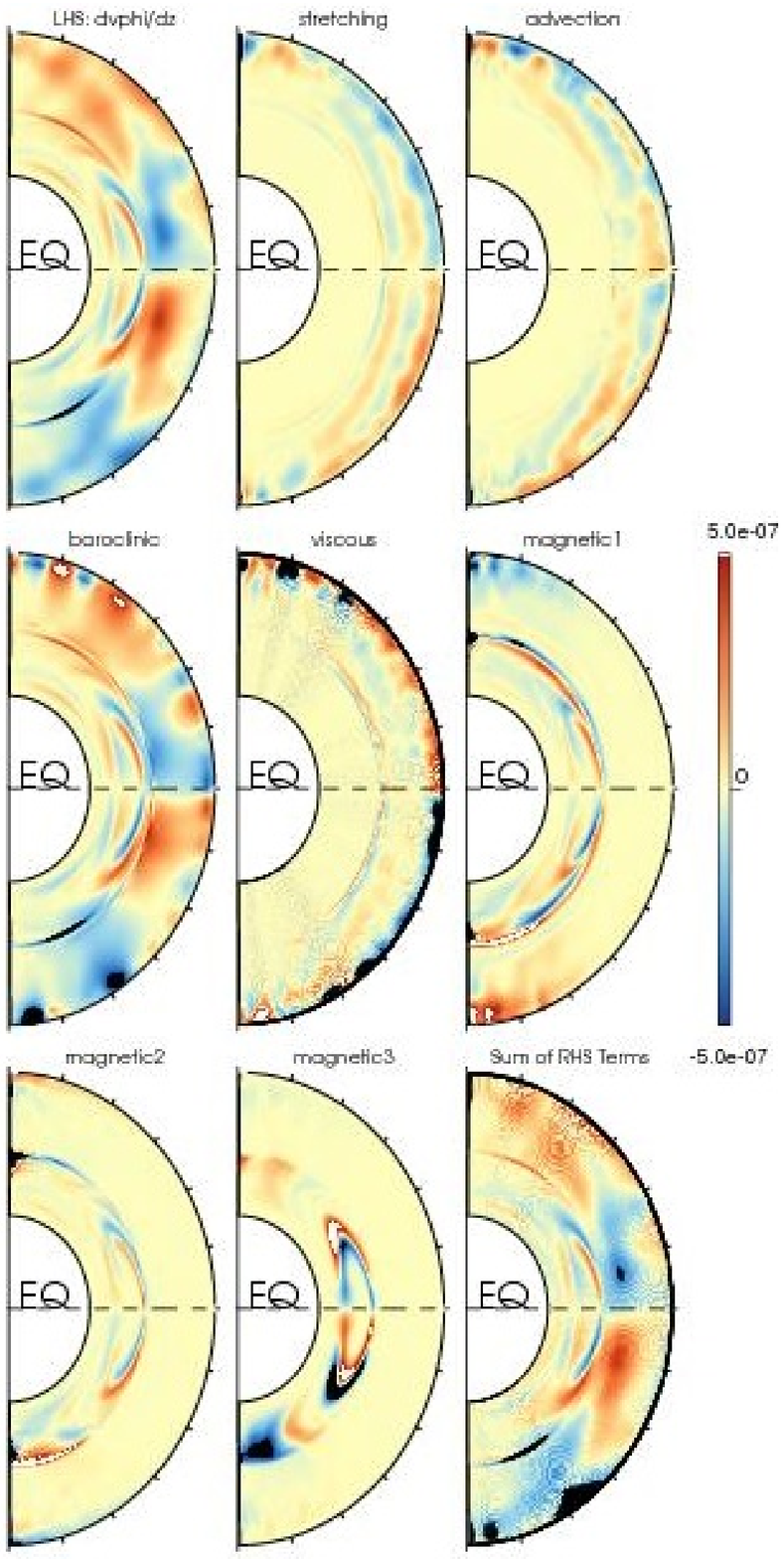}
\caption{Model $M09_{s}$ (anti-solar). Temporal and latitudinal averaged of $\partial v_{\phi}/ \partial z$ (LHS), Reynold stress terms: vortex tube stretching (stretching) and tilting (advection), the baroclinic term in the meridional force balance (baroclinic), the viscous stresses (viscous) and the magnetic terms (magnetic 1,2 and 3; the scale is 100 times smaller for the magnetic terms). The last plot shows the addition of all terms (RHS). The terms are average over 10$\tau_{c}$. Plot units in $s^-1$}.
\end{figure}

\begin{figure}[h]
\centering
\includegraphics[width=1.0\textwidth]{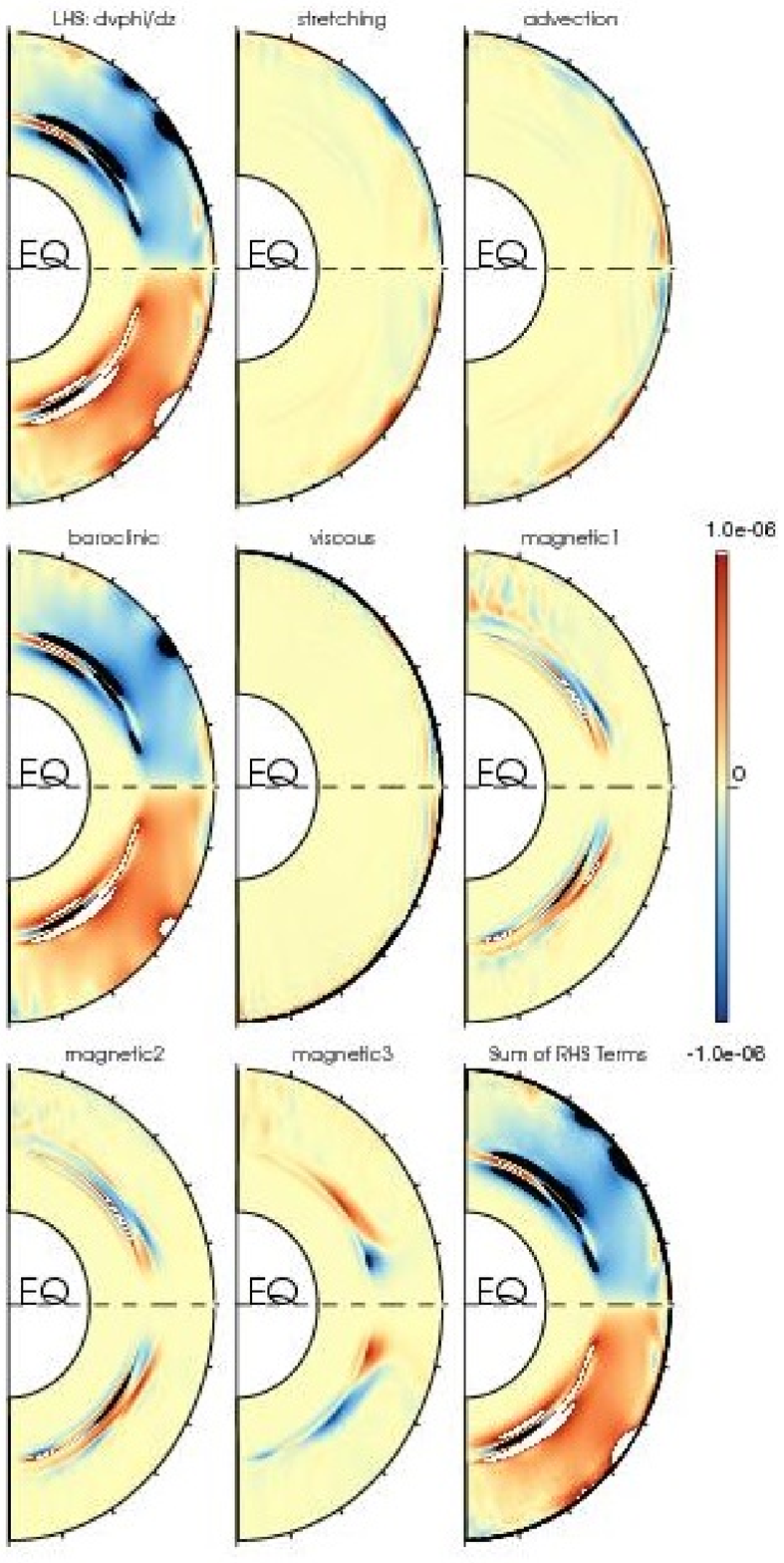}
\caption{Model $M09_{d3}$ (solar). Same than figure 5 (The scale is 50 times smaller for the magnetic terms).}
\end{figure}

To summarize, we calculated the percentage of departure from cylindricity of the differential rotation by the baroclinic term, defined as $|LHS-BAR/LHS|$ where LHS and BAR (baroclinic term) are integrated in radius and latitude. Following thermal wind balance we expect the baroclinic term to increase with rotation (as Fig. 9,10 and 11 confirm), but the baroclinic term becomes relatively less important in the overall balance ($\Delta S \propto \Omega^{n}$ with $0<n<1$) and the difference is mostly compensated by Reynolds stresses. The trend obtained is $|LHS-BAR/LHS| \approx \Omega^{0.26 \pm 0.48}$ in the MHD cases.

\section{Conclusions and dicussion}

The distribution of the energy reservoir in solar-like stars for anti-solar and solar differential rotation cases is different. The main part of the energy in both cases is KE but in the models with solar differential rotation there is a non negligible proportion of ME. We expect this ratio to increase even further as star's rotation increases beyond 3 times the rotation of the Sun reaching equipartition, even super equipartition (as in \citet{2016arXiv160303659A,2015SSRv..196..101B}). The KE in solar differential rotation cases is mainly DRKE (more dominant if the star rotation rate is larger) while in the anti-solar cases it is CKE, pointing out that the convective dynamo in the anti-solar cases is dominated by convective motions, a characteristic of $\alpha^{2}$ dynamos, but in the solar cases the $\Omega$ effect can be large enough to build up $\alpha-\Omega$ or $\alpha^{2}-\Omega$ dynamos. A detailed analysis of the convective dynamos in each model of this article will be carried out in a future publication. 

There are essential differences in the rotation profiles comparing simulations with and without magnetic fields. The presence of magnetic fields drives a speed up of the poles and a slowdown of the equator compared with the hydro simulations. The speed up effect is stronger as the star's rotation rate increases, because the enhancement of the star magnetic field leads to a larger transfer of angular momentum from the equator to the poles. The reason of the poles speed up in the anti-solar cases is due to magnetic field opposing Reynolds stresses.

The trends between the star's differential rotation, rotation rate and mass shows a better agreement with observations for the MHD cases than for the hydro simulations. There is a weaker dependency of the differential rotation with the star rotation rate (there is a drop of the power factor from 0.89 in the hydro cases to 0.44 in the MHD simulations) and star's mass (from 8.68 in the hydro cases to 4.19 in the MHD simulations), as well as flatter profiles if we compare the differential rotation and DRKE with the model Ro. This could explain recent observations of solar-like stars since they are seen to possess magnetic fields. To fully account the effect of the magnetic field in the star’s differential rotation using models with different masses and rotation rates, a systematic analysis of the models parameters is required. In future communications we will refine the trends obtained in the present communication adding new models to the analysis, but the current analysis already shows robust trends.

The radial angular momentum flux balance in the anti-solar differential rotation case show opposite characteristics than the solar one; there is a radially outward transport of angular momentum by Reynolds stresses in the anti-solar case while it is inward in the solar case. The latitudinal angular momentum flux balance in the anti-solar case is dominated by the opposition between convective and meridional circulation motions, but for the solar case there is a more complex interplay between components dominated by the meridional circulation and compensated by viscous and magnetic stresses. 

The baroclinic term in the anti-solar differential rotation case is not the dominant component at high latitudes near the top of the convective layer, the effect of the Reynolds stresses and viscosity are important and should be considered to explain the resulting differential rotation of the model. In the solar case, particularly for the model with the largest rotation rate, the baroclinic term is dominant and the other components are almost negligible in the entire domain with only small contributions very close to the top of the convective layer. Baroclinic effects are stronger as star's rotation increases.

\section*{Acknowledgements}

We have received funding by the European Commission's Seventh Framework Programme under the grant agreement SPACEINN (project number 312844) and ERC STARS2 (207430). We thank S. Matt and O. Do Cao for performing most of the HD models. A. Strugarek is a National Postdoctoral Fellow at the Canadian Institute of Theoretical Astrophysics, and acknowledges support from the Canadas Natural Sciences and Engineering Research Council. We extend our thanks to CNES for PLATO science support and INSU/PNST for our grant. We thank the support of Genci center under grant 1623 and Prace allocation ra1964.

\section*{References}

\bibliography{mybibfile}

\end{document}